\documentclass[]{aastex7}

\usepackage{bm}

\def\JB{\mathrm{Jy~beam^{-1}}}
\def\mJB{\mathrm{mJy~beam^{-1}}}

\def\kms{\mathrm{km~s^{-1}}}
\def\cc{\mathrm{cm^{-3}}}
\def\Msun{M_{\sun}}

\def\Rsun{R_{\sun}}

\graphicspath{{./}{figures/}}

\begin{document}

\title{Inclination Effect on Observational Identification of Outflow Rotation in Magnetohydrodynamics Simulations}

\correspondingauthor{Yusuke Aso}

\author[orcid=0000-0002-8238-7709,sname='Aso',gname='Yusuke']{Yusuke Aso}
\affiliation{Korea Astronomy and Space Science Institute, 776 Daedeok-daero, Yuseong-gu, Daejeon 34055, Republic of Korea}
\affiliation{Division of Astronomy and Space Science, University of Science and Technology, 217 Gajeong-ro, Yuseong-gu, Daejeon 34113, Republic of Korea}
\email[show]{yaso@kasi.re.kr}

\author[orcid=0000-0002-0963-0872,sname='Machida',gname='Masahiro']{Masahiro N. Machida}
\affil{Department of Earth and Planetary Sciences, Faculty of Sciences, Kyushu University, Fukuoka 812-8581, Japan}
\email{machida.masahiro.018@m.kyushu-u.ac.jp}

\begin{abstract}

We investigate the observational signatures of outflow rotation in protostellar systems using magnetohydrodynamics simulations of protostellar evolution with radiative transfer and synthetic observation. The velocity gradient perpendicular to the outflow axis indicates outflow rotation. The rotation signature is clearly seen in the moment 1 map and a position-velocity (PV) diagram across an outflow lobe made from our model with an inclination angle of $i\gtrsim 85\arcdeg$, as in observational studies of protostellar outflows. Velocity projection with lower inclinations distorts the moment 1 map because the outflow vertical (propagation) velocity contributes more to the line-of-sight velocity, leading to an incorrect outflow axis direction. The PV diagram adopting the incorrect outflow axis shows no clear velocity gradient. These effects may prevent us from identifying outflow rotation. Our analysis implies that rotational signatures can be obscured in $\sim 2/3$ to $\sim 4/5$ of the total outflow population ($i<70\arcdeg - 80\arcdeg$), regardless of the evolutionary stage. Complicated structures in observed outflows make it difficult to determine the outflow axis, which may result in the apparent non-detection of outflow rotation.

\end{abstract}




\section{Introduction} \label{sec:intro}

Theoretical studies have predicted the rotation of protostellar outflows, which is considered direct evidence of angular momentum removal from a protostellar system.
\citet{bl.pa82} demonstrated that outflows are magnetocentrifugally driven from the disk when magnetic field lines anchored to the rotating disk are inclined by more than $30\arcdeg$ from the disk’s normal.
The magnetocentrifugally driven outflow efficiently extracts angular momentum from the rotating disk through the magnetic field lines and expels it into interstellar space.

The removal of angular momentum is essential for star formation because a star typically forms from a dense core with a specific angular momentum on the order of $j\sim 10^4~\kms~\mathrm{au}$ \citep{good93}, which would be conserved in the absence of an angular momentum removal mechanism.
If the infalling material reached the solar radius with such a large specific angular momentum, the resulting centrifugal force would be several orders of magnitude stronger than the gravity of a solar-mass star.
This is known as the angular momentum problem.
The magnetocentrifugally driven outflow has been proposed as the most promising solution to this problem \citep{pu.no86}.
Rotating outflows have also been reproduced in numerical simulations of magnetohydrodynamics (MHD) under various conditions \citep[e.g.,][]{mach08, ma.ho13}.


Despite theoretical predictions, observational identification of outflow rotation has been limited to several protostellar systems, even though numerous protostellar outflows have been observed in survey studies \citep[e.g.,][]{ar.sa06, step19}.
Outflow rotation is identified when a velocity gradient perpendicular to the outflow axis is observed over a wide range of spatial extents along the outflow’s major axis.
Such examples include CB 26 \citep{lope23, laun23}, TMC-1A \citep{bjer16}, OMC 2/FIR 6b \citep{mats21}, HH270mms1-A \citep{omur24}, DG Tau B \citep{deva22}, and Orion Source I \citep{hiro17, lope20}.
Not only outflows but also jets show rotation \citep{lee17}.
Note that the outflow axis corresponds to the long axis of the outflow and is parallel to the outflow propagation direction.
This velocity gradient is examined in the intensity-weighted mean velocity (moment 1) map and a position-velocity (PV) diagram along a cut perpendicular to the outflow axis, offset from the midplane (or equatorial disk) of the observed system. 
This distribution of the line-of-sight (rotation) velocity is observed when the outflow axis is nearly parallel to the plane of the sky; otherwise, the vertical (outflow propagation) velocity contributes more to the line-of-sight velocity than the rotational velocity.
Thus, the outflow configuration likely limits the number of outflows in which rotation can be observationally identified.
A quantitative evaluation of this effect is therefore required to understand the discrepancy between the theoretical ubiquity and the observational scarcity of outflow rotation.
Such an evaluation will also help assess whether outflow rotation serves as a viable solution to the angular momentum problem.


This paper is structured as follows.
Section~\ref{sec:settings} describes the setup of the MHD simulation used in this study, along with the parameters for radiative transfer and synthetic observations applied to the simulation results.
In Section~\ref{sec:anares}, we present a simulation snapshot with physical properties similar to those of an observed rotating outflow, as well as the moment maps and PV diagrams derived from the snapshot.
Section~\ref{sec:discussion} discusses how the combination of outflow propagation and rotational velocities affects the line-of-sight velocity gradient and its implications for the detection of outflow rotation.
Finally, we summarize our findings in Section~\ref{sec:conc}.

\section{Numerical Settings} \label{sec:settings}
\subsection{Magnetohydrodynamics Simulation} \label{sec:mhd}
The numerical simulation used in this study is the same as that in \citet{tomi17}, \citet{as.ma20}, and \citet{ma.ba24}.
A brief summary is provided below.

The simulation begins with a critical Bonnor-Ebert sphere.
The initial conditions are as follows: a central molecular hydrogen density of $6\times 10^5~\cc$, an isothermal temperature of 10\,K, a radius of 0.03\,pc, a mass of $1.25~\Msun$, a uniform magnetic field of $51~\mu \mathrm{G}$ in the $z$-direction, and an angular rotation frequency of $2\times 10^{-13}~\mathrm{s}^{-1}$.
The initial rotational axis is aligned with the magnetic field direction. 

We use a nested grid code with 13 levels of refinement ($l = 1$-$13$, where $l$ denotes the grid level) \citep[for details, see][]{Machida2004,Machida2014,Machida2016}.
Each grid consists of $64\times 64 \times 32$ cells, adopting mirror symmetry across the equatorial plane in the $z$-direction, with cell sizes $h(l)$ ranging from $h(13) = 0.75$\,au to $h(1) = 3300$\,au.
The outer boundary is fixed at a radius of 0.5~pc, which is 16 times larger than the initial core radius, while the inner boundary is a sink cell with a size of 1~au.
This simulation solves the equations of resistive MHD and employs a barotropic equation of state \citep{Machida2006} to calculate the evolution of the collapsing core until the protostellar mass reaches $M_*\sim 0.5\Msun$.

\subsection{Radiative Transfer and Synthetic Observation}
As described above, we adopt the barotropic equation of state without considering heating from the protostar, which can result in lower temperatures in low-density regions, such as the outflow lobe.
Thus, this approach might not accurately reflect the thermal structure expected when radiative heating from the central protostar is considered.
To address this issue, following \citet{tomi17}, we calculate the temperature field using the open-source code RADMC-3D \citep{dull12}, adopting a typical protostellar radius of $2~\Rsun$, a typical protostellar temperature of 4000~K, and the default dust model of RADMC-3D (silicate) to obtain a more realistic temperature distribution than that based on the barotropic equation of state.

To produce observable quantities under conditions similar to actual observations, we refer to a rotating outflow observed in the protostellar system CB 26, which has an inclination angle of $i=85\arcdeg$, as presented in \citet{lope23}.
To calculate the intensity, radiative transfer simulations are performed using RADMC-3D with the density and velocity fields obtained from the numerical simulation (see Section~\ref{sec:mhd}) and the recalculated temperature field.
The intensity is computed for the CO $J=2-1$ line, which was used in \citet{lope23} to detect the outflow, assuming a CO abundance of $2.7\times 10^{-4}$ relative to H$_2$ molecules \citep{lacy94}.
In addition, the dust continuum intensity is calculated in the same frequency channels as the line emission for continuum subtraction, adopting a gas-to-dust mass ratio of 100.
The pixel size and image size are set to $0\farcs 1$ and $50\arcsec$, respectively.
The velocity channels have a resolution of $0.1~\kms$ and cover a range of $-10$ to $10~\kms$.
The intensity is scaled based on the distance to CB 26, which is 140~pc.

The interferometric effect is incorporated using the Common Astronomy Software Applications \citep[CASA;][]{casa22}. 
The visibilities are computed from the continuum-subtracted data cube obtained above using the CASA task {\it simobserve}. 
A single ALMA antenna configuration, Cycle-11 C43-5, is adopted to achieve an angular resolution similar to that of \citet{lope23} ($1\farcs 2 \times 0\farcs 9$). 
The input model image is centered at the coordinates of CB 26, $(\alpha_\mathrm{J2000}, \delta_\mathrm{J2000}) = (04^\mathrm{h}59^\mathrm{m}50\fs 742, 52\arcdeg 04\arcmin 43\farcs 49)$. The total observation time and integration time are set to 180\,s and 60\,s, respectively, corresponding to three scans during the synthetic observation.

After {\it simobserve}, artificial Gaussian noise is added to each visibility using the CASA task {\it setnoise} with the {\it simplenoise} argument. 
The obtained visibilities are Fourier transformed using the CASA task {\it tclean}. The pixel and image sizes are the same as those of the input cube in {\it simobserve}. 
The velocity resolution is changed to $0.18~\kms$, matching the value used in \citet{lope23}. 
The deconvolver is set to `multiscale' with scales of 0, 5, 10, and 20 pixels.

The CLEAN process employs Briggs weighting with a robust parameter of 0.5. The clean components are obtained down to $30~\mJB$. A uv-taper of [0.8 arcsec, 0.01 arcsec, 90 deg] is applied. The resulting angular resolution and noise level in emission-free channels are $1\farcs 25 \times 0\farcs 88$ and $10~\mJB$, respectively, closely matching those of \citet{lope23}.

\section{Analysis and Results} \label{sec:anares}
\subsection{Snapshot for CB 26}
\begin{figure}[htbp]
\gridline{
\fig{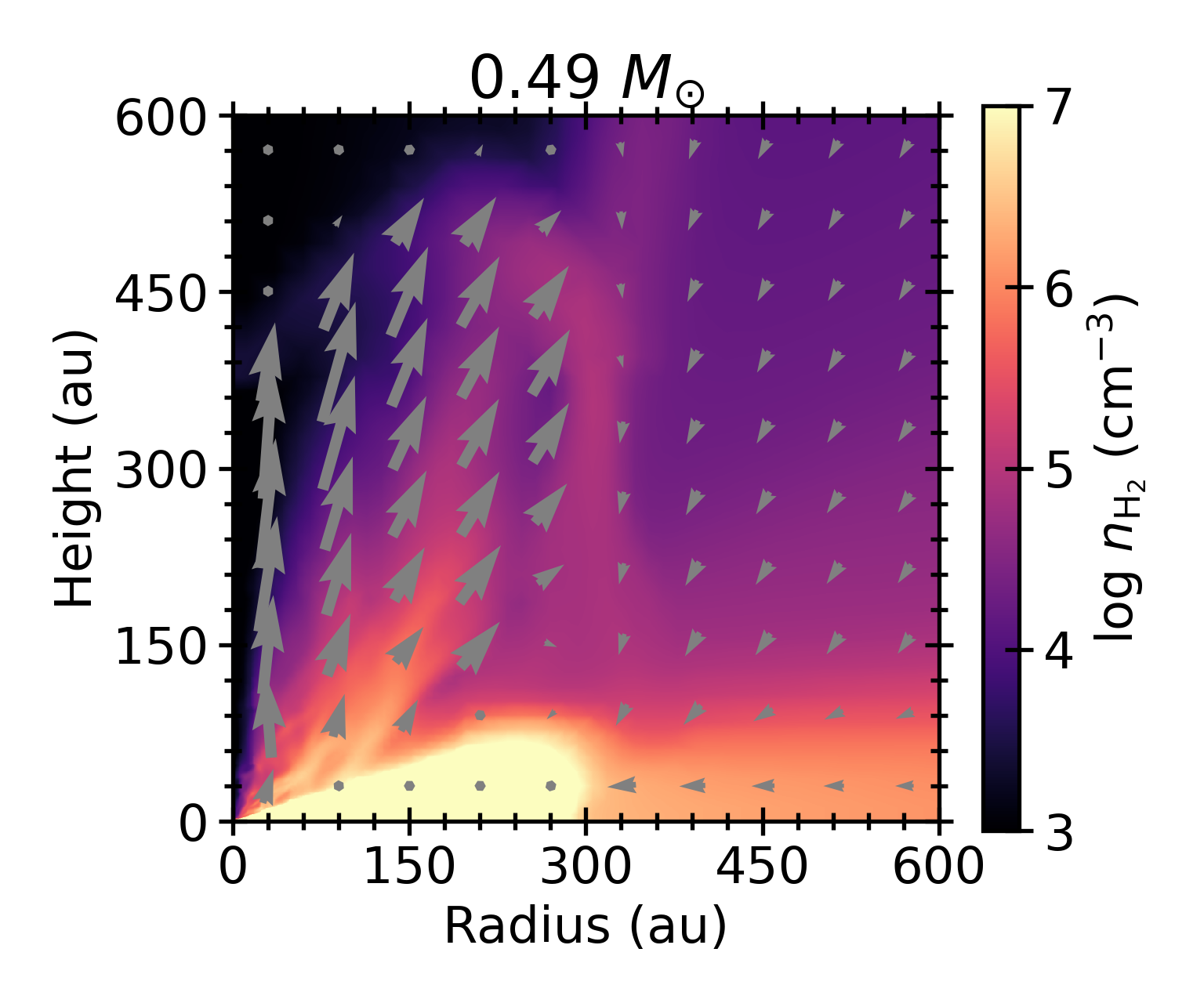}{0.49\textwidth}{(a)}
\fig{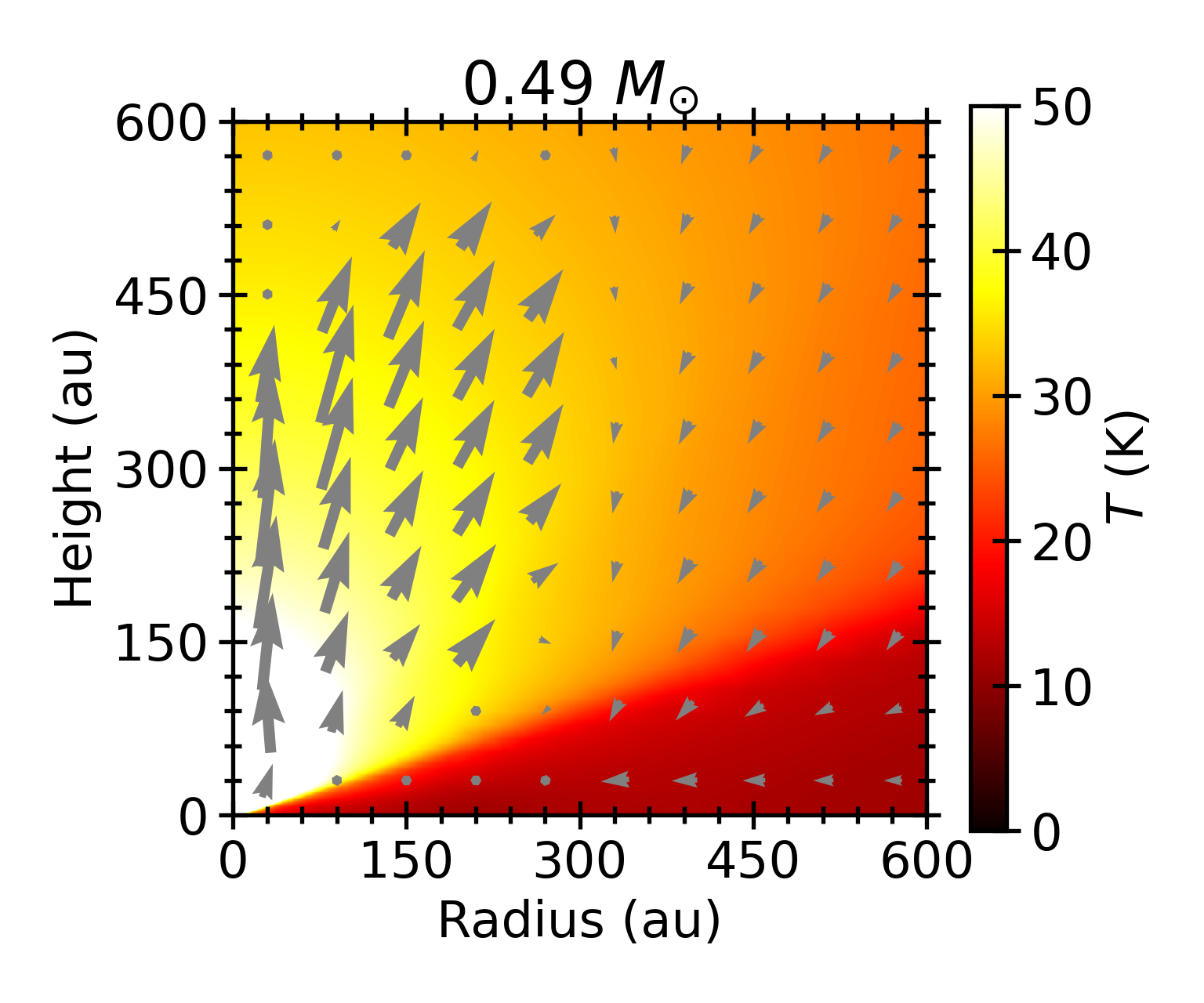}{0.49\textwidth}{(b)}
}
\gridline{
\fig{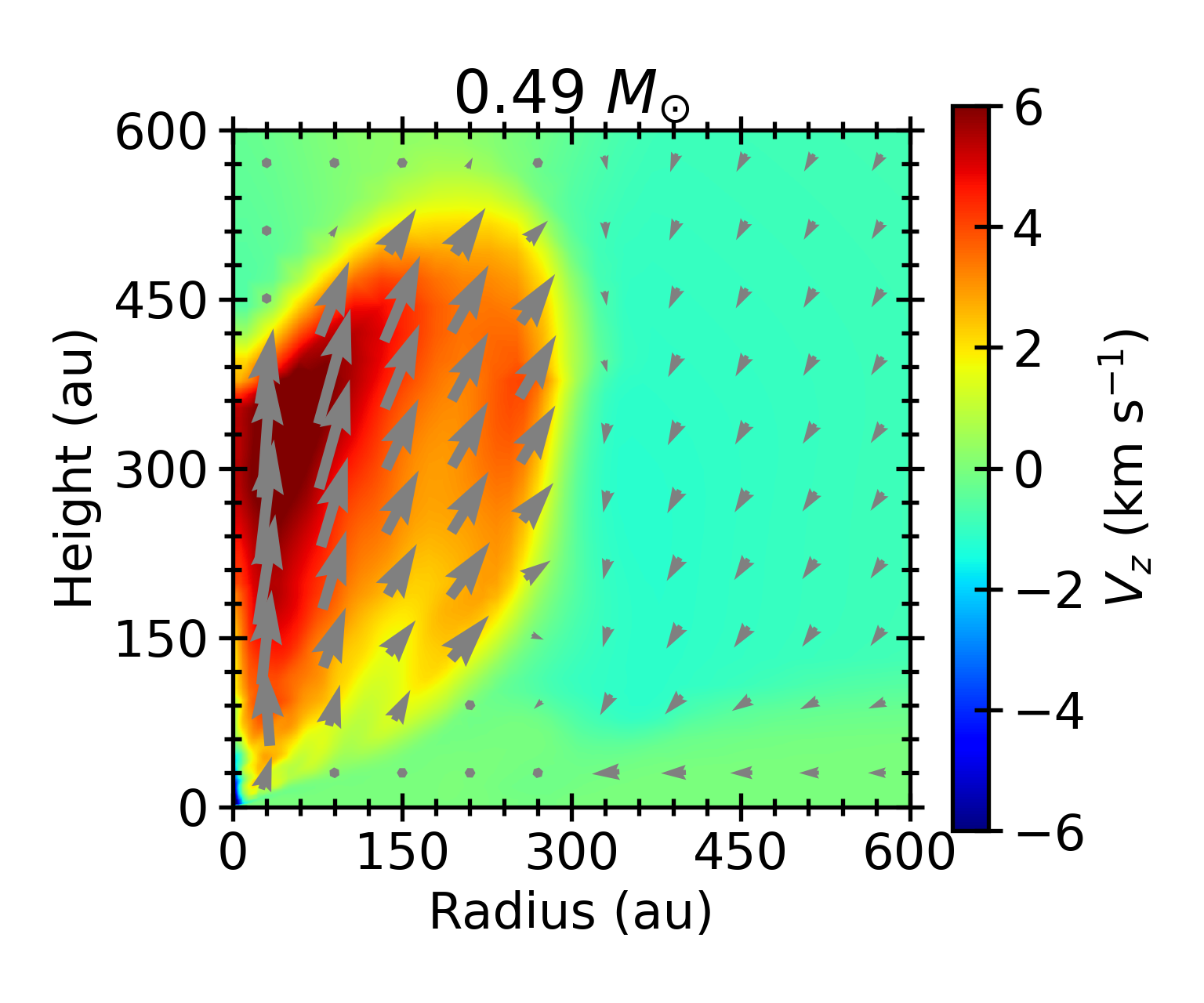}{0.49\textwidth}{(c)}
\fig{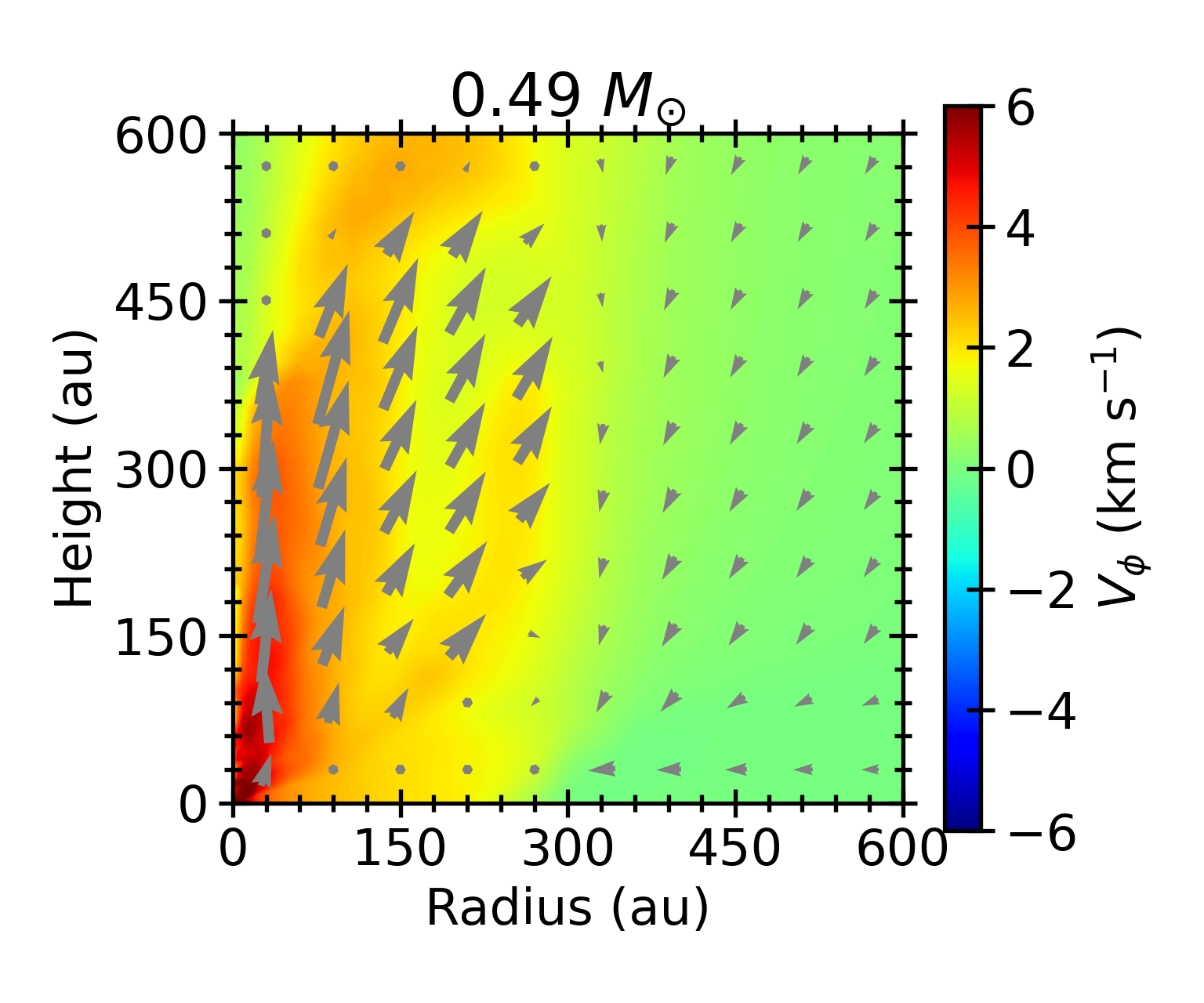}{0.49\textwidth}{(d)}
}
\caption{
Azimuthally averaged physical quantities at the snapshot for CB 26, plotted at the same cylindrical radius: (a) number density, (b) temperature, (c) vertical velocity, and (d) azimuthal (rotational) velocity. 
The arrows indicate the direction and magnitude of the cylindrical radial velocity and vertical velocity.
\label{fig:phys}
}
\end{figure}

Our MHD simulation provides snapshots with physical properties similar to those of the observed rotating outflow in the protostellar system CB 26.
Figure~\ref{fig:phys} presents azimuthally averaged physical quantities taken from a snapshot in which the protostellar mass is $M_*=0.49~\Msun$.
For reference, the central stellar mass of CB 26 is estimated to be $0.66~\Msun$ in \citet{lope23}.
Figure~\ref{fig:phys} shows the hydrogen number density $n_\mathrm{H_2}$, temperature $T$, vertical velocity $V_z$ (associated with outflow propagation), and azimuthal (rotational) velocity $V_\phi$.
The hydrogen number density is derived from the MHD simulation, where a mean molecular weight of 2.8 is adopted.

The arrows indicate an outflow lobe with a length of $\sim 540$\,au and a width (or cylindrical radius) of $\sim 270$\,au.
This length is comparable to that of the CB 26 outflow \citep{lope23}.
For this reason, we focus on this 540-au-sized lobe, although the entire outflow extends to 15,000\,au at this epoch \citep{ma.ba24}.
The hydrogen number density is $\lesssim 10^3\,\cc$ in the inner regions of the lobe, while it reaches $\sim 3\times 10^5~\cc$ in the outer regions.
Inside the outflow lobe, the temperature is typically $\sim 40$~K and generally decreases with increasing spherical radius, as expected from heating by stellar radiation. This order of temperatures is also consistent with estimation in observational studies of protostellar outflows \citep{gome19}.

The total mass of this outflow lobe is calculated by integrating the gas mass over the region where the azimuthally averaged $V_z > 1~\kms$ within $r < 600$\,au.
The resulting outflow mass is $1.5\times 10^{-4}\,\Msun$.
Although this value is three times larger than the mass of the CB 26 outflow reported in \citet{lope23}, their estimate should be considered a lower limit since it is derived from the CO $J=2-1$ line, which is likely optically thick.
In addition, this difference could be due to unknown cloud properties, such as CO abundance and density.

The highest azimuthally averaged vertical velocity in this snapshot is $V_z \sim 7~\kms$, appearing in the inner region of the outflow, while the outer part of the lobe has a typical vertical velocity of $V_z \sim 3~\kms$.
The rotational velocity $V_\phi$ is also higher in the inner regions ($\sim 4~\kms$) than in the outer regions ($\sim 2~\kms$).
This gradient of rotational velocity is also reported in observational studies \citep{zapa10}.
In addition, the rotational velocity in some parts of the lobe exceeds that in the disk midplane.
Since the outflowing gas, which is launched from the region near the disk surface, is accelerated by the magnetocentrifugal driving mechanism, the rotational velocity above the disk should be higher than that on the disk surface \citep{mats21}.
Table~\ref{tab:comp} summarizes the comparison of the stellar mass $M_*$, outflow mass $M_\mathrm{flow}$, outflow length $l_\mathrm{flow}$, vertical velocity $V_z$, and rotational velocity $V_\phi$ between the outflows of CB 26 and our model.

\begin{deluxetable}{cccccc}
\tablecaption{Comparison of outflow properties between CB 26 and our model. \label{tab:comp}}
\tablehead{
 & $M_*~(\Msun)$ & $M_\mathrm{flow}~(10^{-4}~\Msun)$ & $l_\mathrm{flow}$~(au) & $V_\mathrm{z,max}~(\kms)$ & $V_\phi~(\kms)$
}
\startdata
CB 26 & 0.66 & $>0.5$ & $\sim 540$ & 10 & 1-3\\
Model & 0.49 & 1.5 & $\sim 540$ & 7 & 2-4
\enddata
\tablecomments{
The physical quantities of CB 26 are taken from \citet{lope23}.
The outflow mass $M_\mathrm{flow}$ includes both the upper and lower lobes, while the outflow length $l_\mathrm{flow}$ refers to the length of each lobe.
The outflow mass $M_\mathrm{flow}$ for CB 26 should represent a lower limit since the CO line is likely optically thick.
The $M_\mathrm{flow}$ and $l_\mathrm{flow}$ in our model are derived from the outflow within a 600\,au $\times$ 600\,au region, considering only gas with an azimuthally averaged vertical velocity exceeding $V_z > 1~\kms$.
}
\end{deluxetable}

\subsection{Observable Signatures of Outflow Rotation} \label{sec:sign}
To verify whether outflow rotation can be observationally identified, we generated moment 0 and moment 1 maps, as well as PV diagrams, from the image cubes of the CB 26 snapshot, both with and without the interferometric effect, after performing the radiative transfer calculation.
Moment 0 represents the intensity integrated over velocity channels, while moment 1 represents the intensity-weighted mean velocity.
These maps are commonly used to identify outflow rotation in observational studies \citep[e.g.,][]{hiro17, lope20, lope23, mats21}.

\begin{figure}[htbp]
\gridline{
\fig{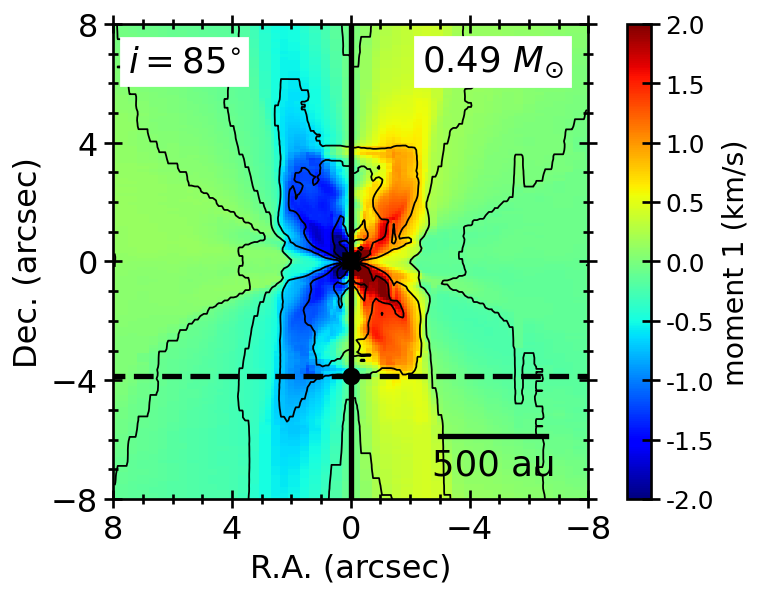}{0.40\textwidth}{(a) Without interferometric effect.}
\fig{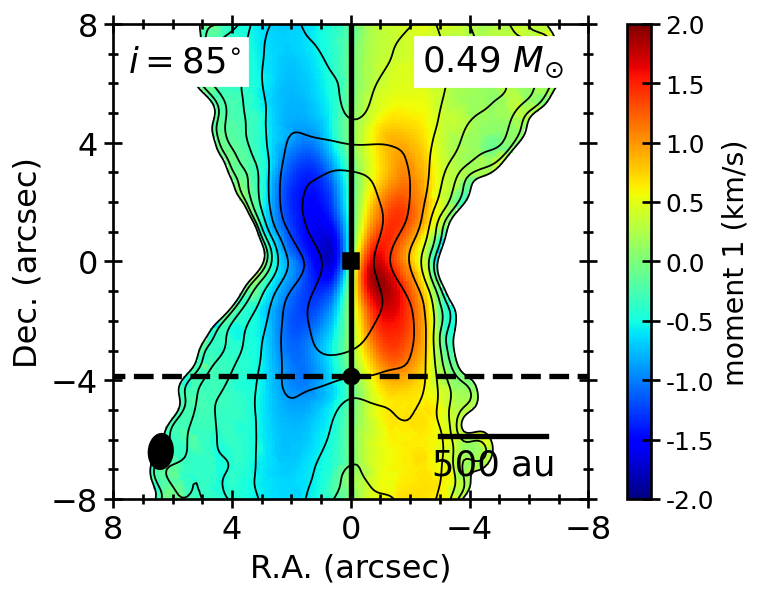}{0.40\textwidth}{(b) With interferometric effect.}
}
\caption{
Moment 0 and moment 1 maps of the CB 26 snapshot with an inclination angle of $i=85\arcdeg$. 
The contours represent moment 0 (integrated intensity), while the color scale indicates moment 1 (intensity-weighted mean velocity). 
The intensities are integrated over the velocity range from $-5$ to $5~\kms$. The left and right panels are generated from the data cube without and with the interferometric effect, respectively. 
The contour levels are $3,6,12,24...\times 0.5~\mathrm{mJy~pixel^{-1}}$ and $13.6\mJB$ in panels (a) and (b), respectively. 
In panel (b), regions where the moment 0 intensity falls below the $3\sigma$ level of $40.8~\mJB$ are masked out.
\label{fig:fidmom}}
\end{figure}

Figure~\ref{fig:fidmom} shows the moment 0 and 1 maps of the snapshot for the CB 26 model with the inclination angle of $i=85\arcdeg$ in which the inclination angle is the same as the observation of CB 26 outflow \citep{lope23}. 
The intensity of the image cubes is integrated from $-5$ to $5~\kms$ to make the moment maps. 
This range covers all the emission from the outflow. 
These moment 0 maps (contours) clearly trace the outflow in the vertical direction with a neck at the declination offset of zero (i.e., midplane). 
The moment 1 (color scale) shows a clear velocity gradient from the left (blueshifted velocity) to the right (redshifted velocity) at any declination, which is caused by the rotation of the outflow. These morphology and velocity gradients are seen in both cases with and without the interferometric effect. These results are consistent with the observational results of the CB 26 outflow \citep{lope23}, demonstrating that the moment maps can be used to identify outflow rotation with the high inclination angle.

\begin{figure}[htbp]
\gridline{
\fig{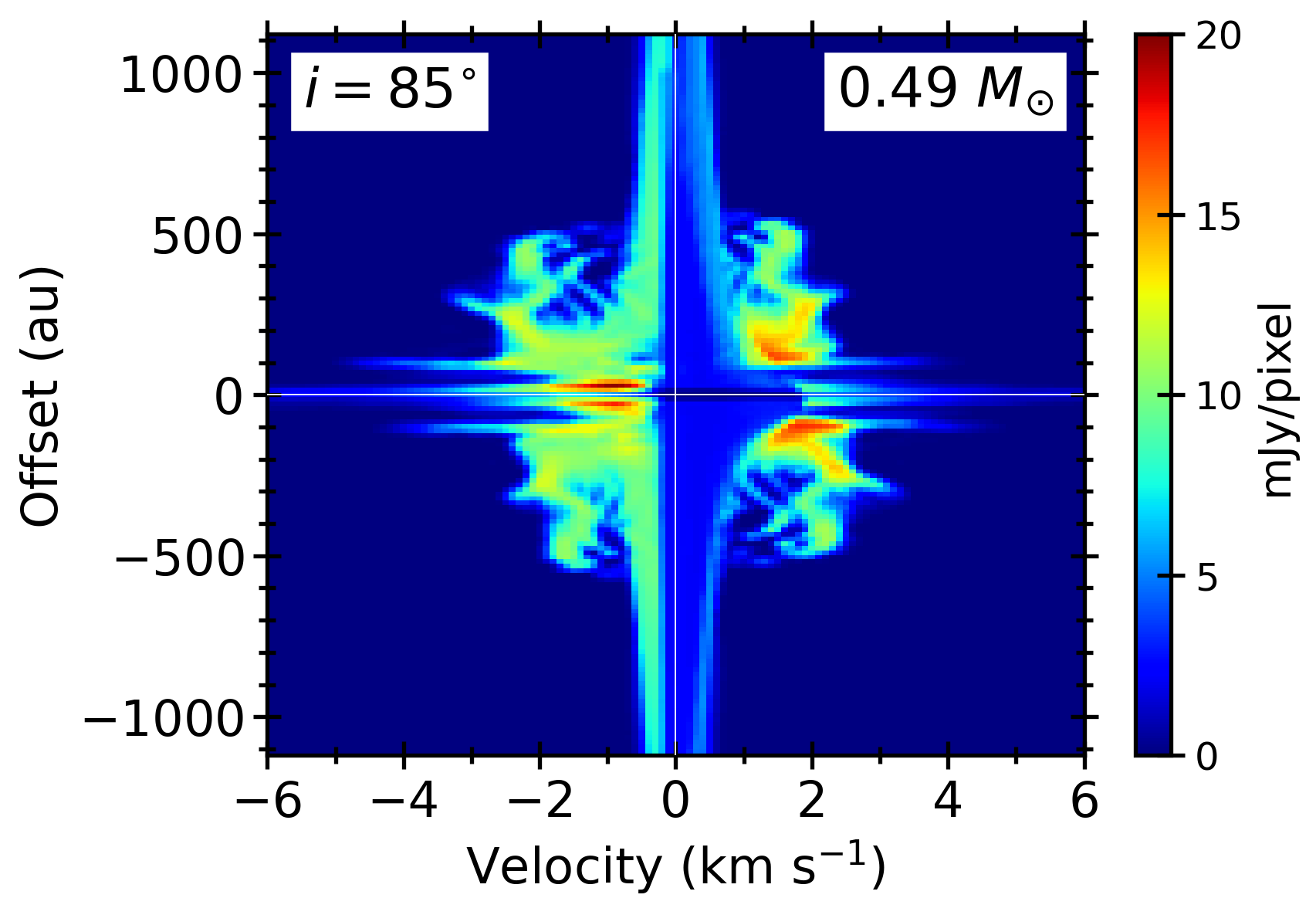}{0.42\textwidth}{(a) Along the black solid line in Figure \ref{fig:fidmom}(a).}
\fig{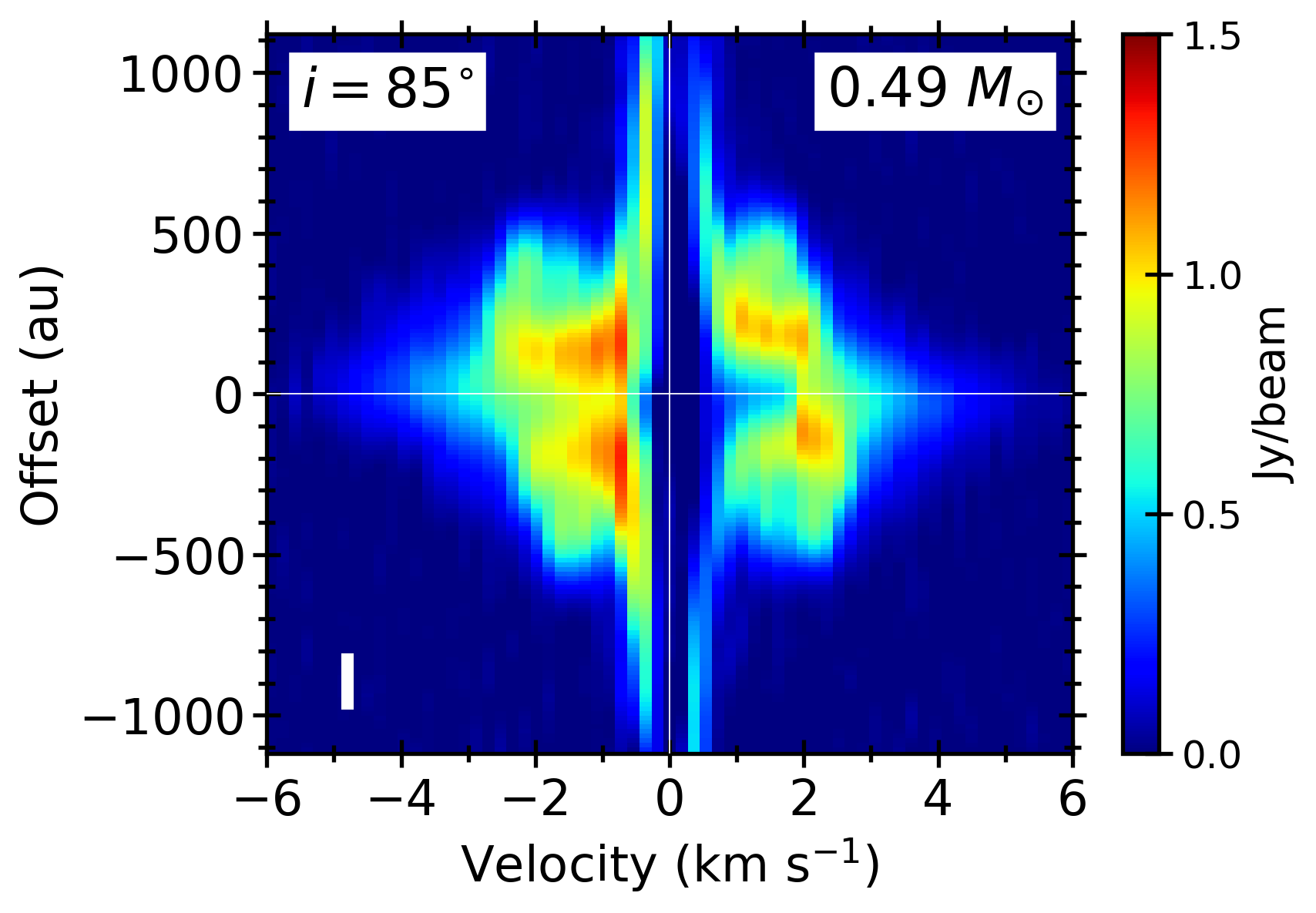}{0.42\textwidth}{(b) Along the black solid line in Figure \ref{fig:fidmom}(b).}
}
\gridline{
\fig{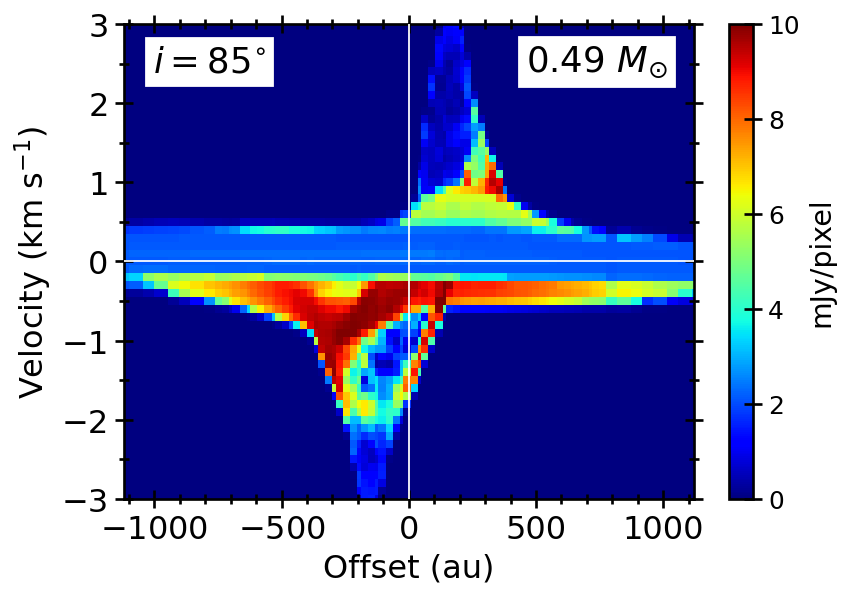}{0.38\textwidth}{(c) Along the black dashed line in Figure \ref{fig:fidmom}(a).}
\fig{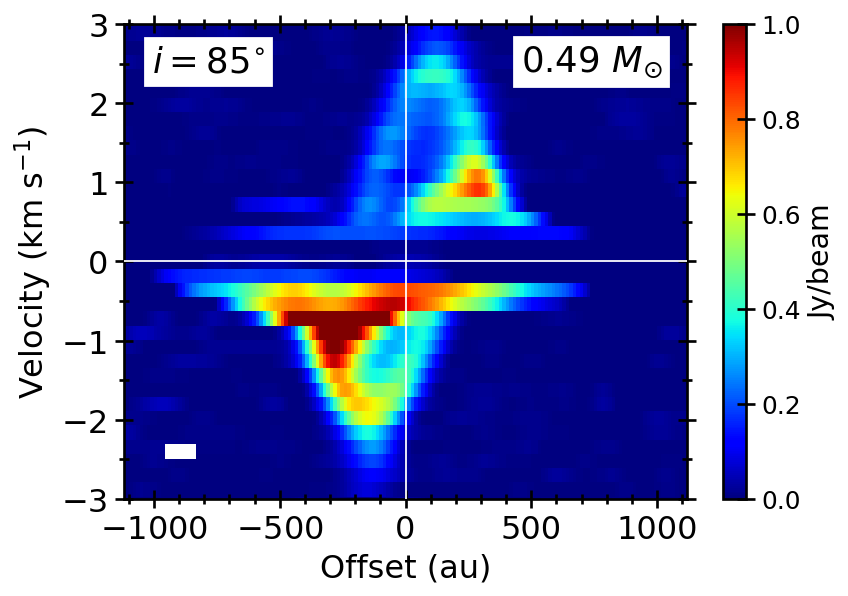}{0.38\textwidth}{(d) Along the black dashed line in Figure \ref{fig:fidmom}(b).}
}
\caption{
Position-velocity diagrams generated from the CB 26 snapshot. 
Panels (a) and (b) show cuts along the outflow axis, while panels (c) and (d) show cuts along a line perpendicular to the outflow axis, offset by 540~au $\times \sin i$ from the protostellar position. 
The white rectangles in panels (b) and (d) indicate the spatial and velocity resolutions of the synthetic observation.
\label{fig:fidpv}}
\end{figure}

Figure~\ref{fig:fidpv} shows PV diagrams along the outflow axis (panels (a) and (b)) and along a cut perpendicular to the outflow axis (panels (d) and (d)), offset by 540 au $\times \sin i$ in the declination direction.
The 540-au offset is also adopted for the CB 26 outflow in \citet{lope23}.
The PV diagram along the outflow axis consists of three components: a vertically elongated component ($>2000$\,au) near zero velocity (i.e., the systemic velocity), a widely distributed high-velocity component ($\sim 10\,\kms$) near the stellar position, and a third component roughly spanning the region of $[-600~\mathrm{au}, 600~\mathrm{au}] \times [-3~\kms, 3~\kms]$ in both cases—without (Figure~\ref{fig:fidpv}(a)) and with (Figure~\ref{fig:fidpv}(b)) the interferometric effect.

The third component traces the outflow lobe, as its spatial and velocity distributions are consistent with those of the outflow in Figure~\ref{fig:phys}.
This component extends over the region of $[-600~\mathrm{au}, 600~\mathrm{au}] \times [-3~\kms, 3~\kms]$. This is mainly because the cylindrically radial component of the outflow velocity causes both blue- and redshifted line-of-sight velocity at any point along the outflow. In addition, the overall structure exhibits a slight velocity gradient from the positive offset (blueshifted velocity) to the negative offset (redshifted velocity).
This velocity gradient is attributed to the propagation (vertical) velocity ($V_z$) of the outflow.


The PV diagrams along the line perpendicular to the outflow with the declination offset exhibit a clear velocity gradient from left (blueshifted velocity) to right (redshifted velocity) in both cases without (Figure \ref{fig:fidpv}(c)) and with (Figure \ref{fig:fidpv}(d)) the interferometric effect. 
This velocity gradient is also evident in the moment 1 map (Figure \ref{fig:fidmom}) and is attributed to outflow rotation. 
More specifically, the PV diagram without the interferometric effect reveals higher velocities at positions closer to the center. 
This feature supports that the PV diagram traces outflow rotation, because the outflow lobe exhibits faster rotation at smaller radii (Figure \ref{fig:phys}(d)). The highest-velocity ($<-2~\kms$ and $>1.5~\kms$) components at smaller offsets ($-200$ to $200$~au) appear faintest in this PV diagram, which can be  attributed to the density distribution inside the outflow lobe (Figure \ref{fig:phys}(a)). 
In comparison, these components appear more clearly in the PV diagram with the interferometric effect, forming an elongated ring or parallelogram shape. 
The weaker emission enclosed by this ring-like component suggests that the line-of-sight velocity is nonzero at the stellar position. 
This is because the line-of-sight velocity includes not only rotational motion but also expansion in the cylindrically radial direction. 
The velocity gradient in the PV diagram across an outflow lobe is consistent with those observed in the CB 26 outflow \citep{lope23}. 
In addition, the velocity range, $|V|\lesssim 3~\kms$, and the peak intensity, $\sim 1~\JB$, align well with those in the PV diagram of CB 26.

\subsection{Effects of Inclination Angles} \label{sec:effects}
In the previous subsection, we confirmed that outflow rotation can be observationally identified under the physical and observational conditions of the CB 26 outflow.
In this subsection, we investigate the effect of the inclination angle on this identification.
Since the inclination angle determines the contribution of rotational velocity to the line-of-sight velocity, it is expected to significantly influence the observed velocity structure.

\begin{figure}[htbp]
\gridline{
\fig{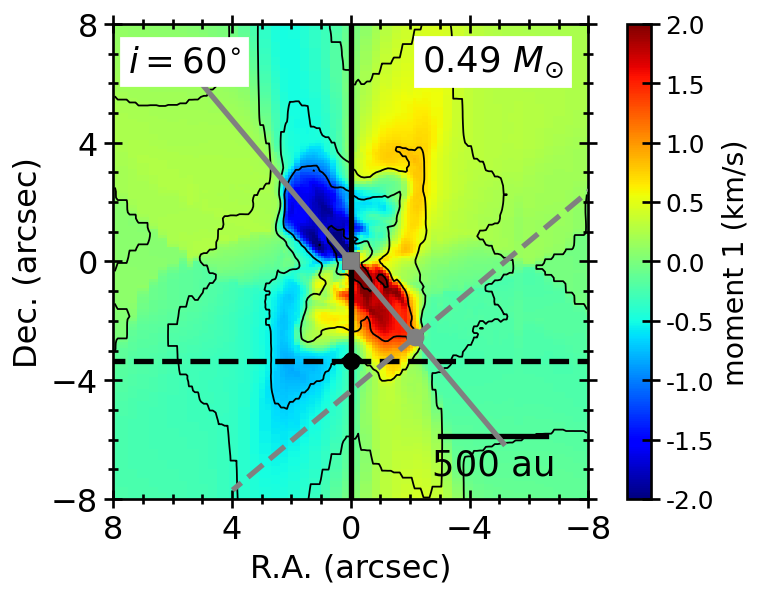}{0.40\textwidth}{(a) Without interferometric effect.}
\fig{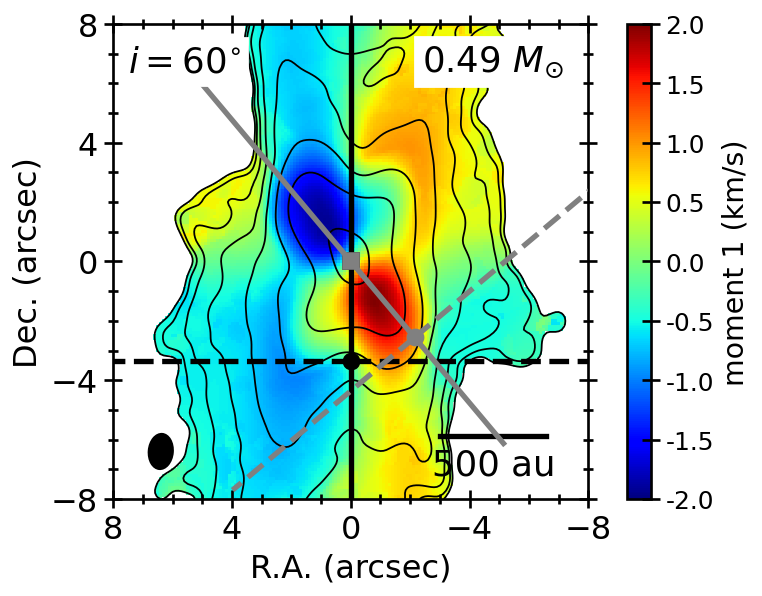}{0.40\textwidth}{(b) With interferometric effect.}
}
\caption{
Same as Figure \ref{fig:fidmom}, but with an inclination angle of $60\arcdeg$.
The gray solid and dashed lines are rotated counterclockwise by $40\arcdeg$ around the center relative to the black solid and dashed lines.
If the gray solid line is misinterpreted as the outflow axis, the gray pair is used instead of the black pair.
\label{fig:60mom}
}
\end{figure}

Figure \ref{fig:60mom} presents the moment 0 and moment 1 maps of the CB 26 snapshot with an inclination angle of $i=60\arcdeg$, which is lower than that of the observed CB 26 outflow.
This figure gives a markedly different impression compared to the case with $i=85\arcdeg$: the most prominent velocity gradient extends from the highest blueshifted velocity in the upper left to the highest redshifted velocity in the lower right, at a position angle of $\sim 40\arcdeg$ (solid gray line in Figure \ref{fig:60mom}).
The second most prominent velocity gradient appears from the lower left (blueshifted) to the upper right (redshifted).
Both velocity gradients are present in cases with and without the interferometric effect and correspond to rotation of the outflow (or the entire system), where the blueshifted velocity is on the left and the redshifted velocity is on the right.

However, the identification of rotation may depend on the assumed direction of the outflow axis.
The most prominent velocity gradient makes the outflow direction less clear in the moment maps with $i=60\arcdeg$ (Figure \ref{fig:60mom}) than in those with $i=85\arcdeg$ (Figure \ref{fig:fidmom}).
In addition, unlike the case with $i=85\arcdeg$ (Figure \ref{fig:fidmom}), the moment 0 map for $i=60\arcdeg$ (Figure \ref{fig:60mom}) lacks the characteristic neck structure.
This further obscures the outflow direction in the moment 0 map.

\begin{figure}[htbp]
\gridline{
\fig{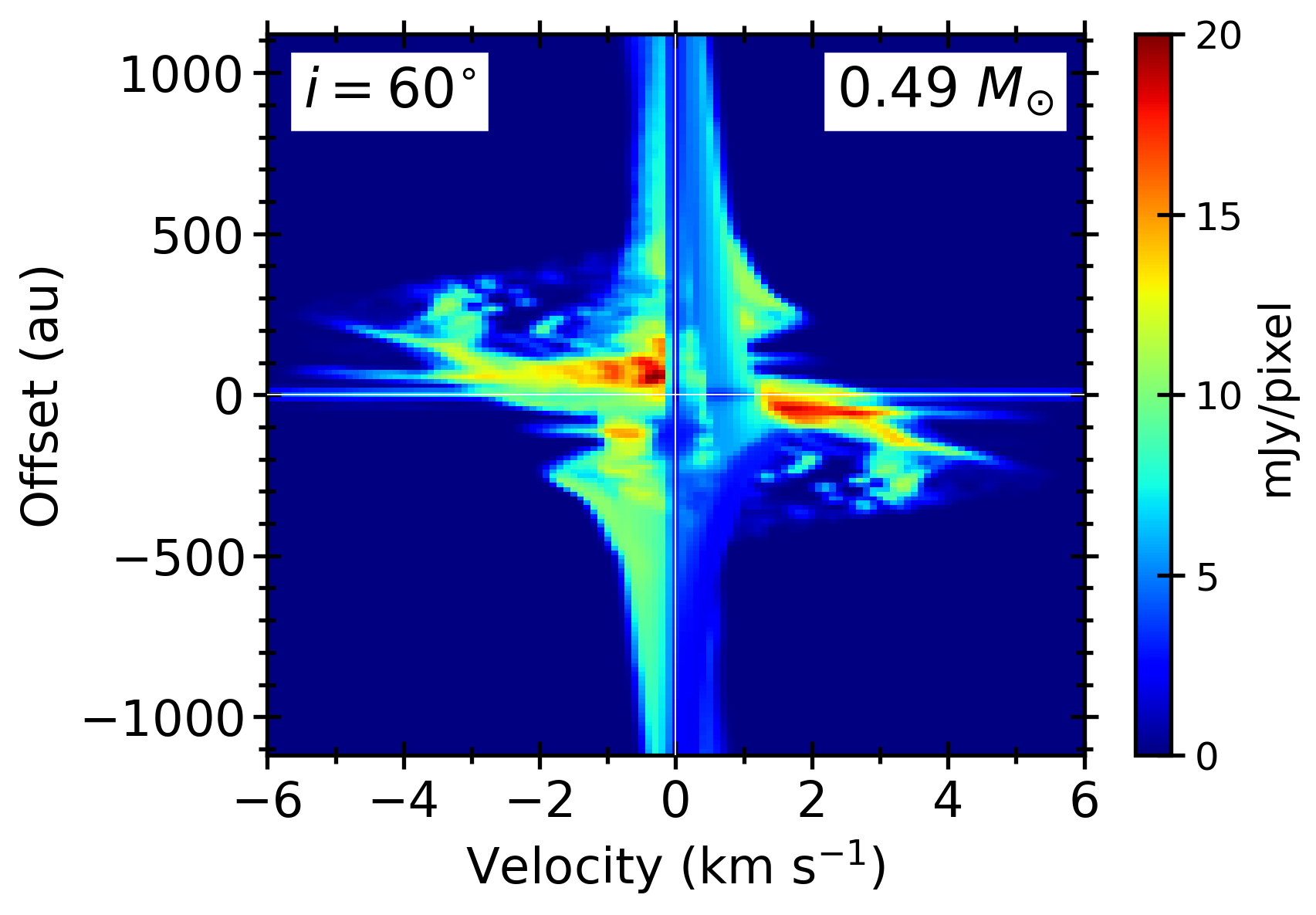}{0.42\textwidth}{(a) Along the black solid line in Figure \ref{fig:60mom}(a).}
\fig{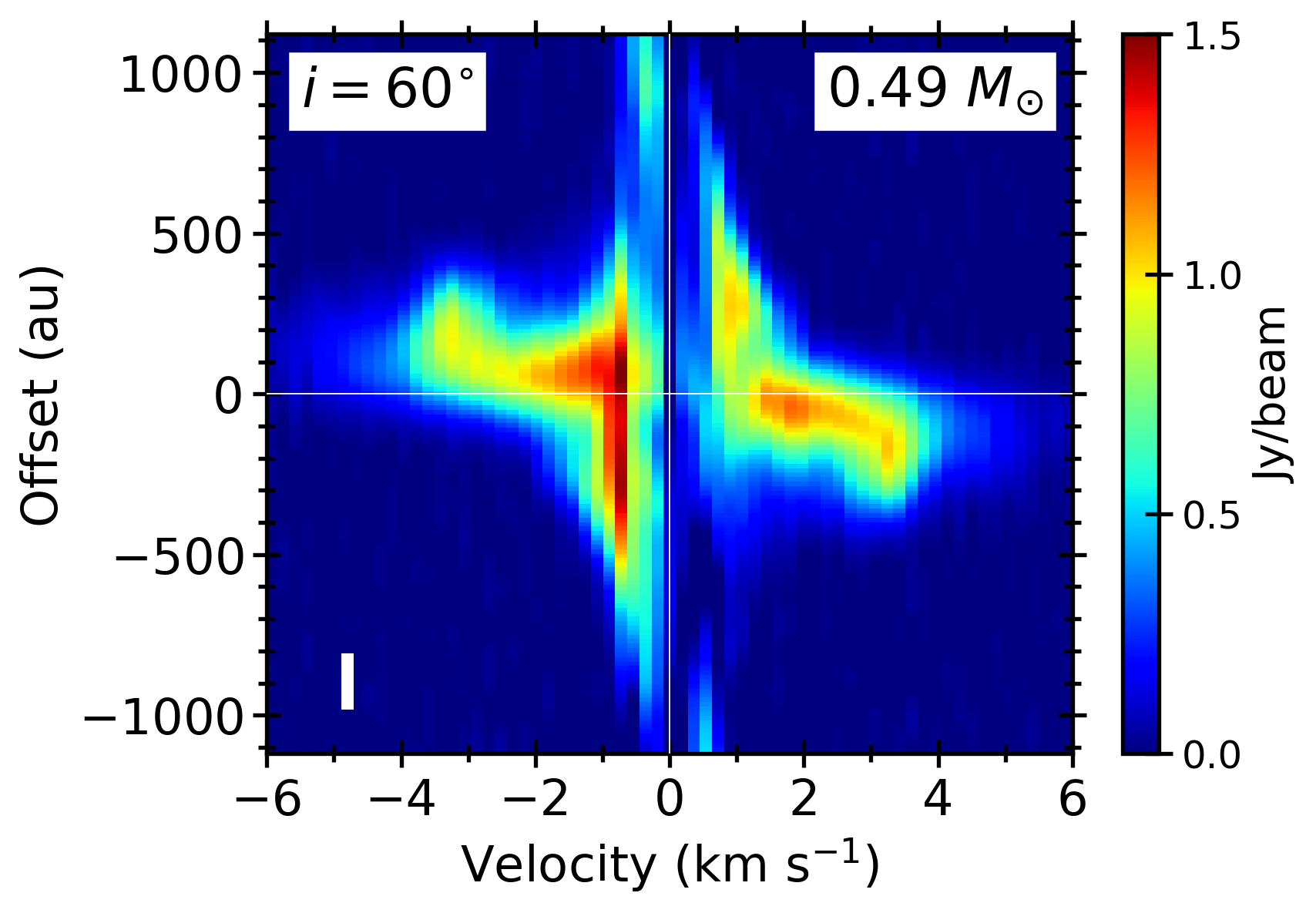}{0.42\textwidth}{(b) Along the black solid line in Figure \ref{fig:60mom}(b).}
}
\gridline{
\fig{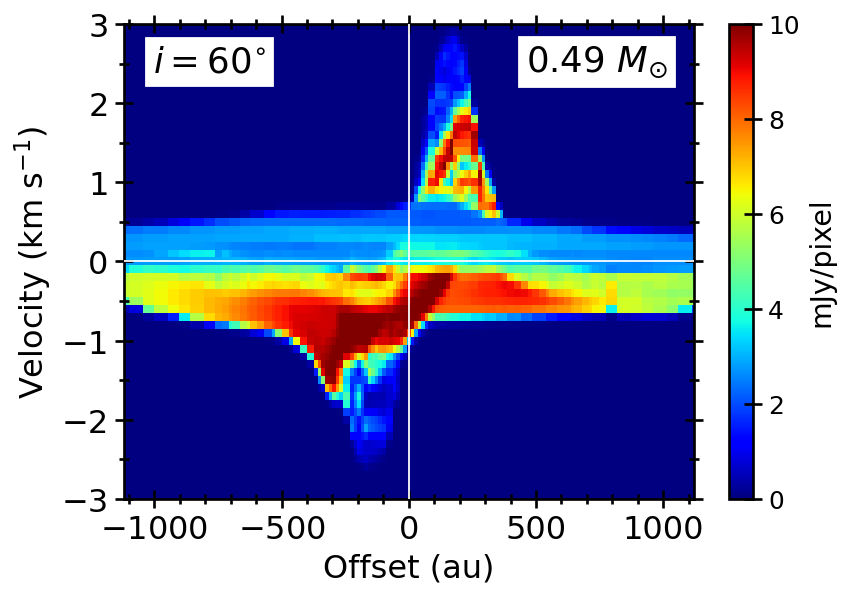}{0.38\textwidth}{(c) Along the black dashed line in Figure \ref{fig:60mom}(a).}
\fig{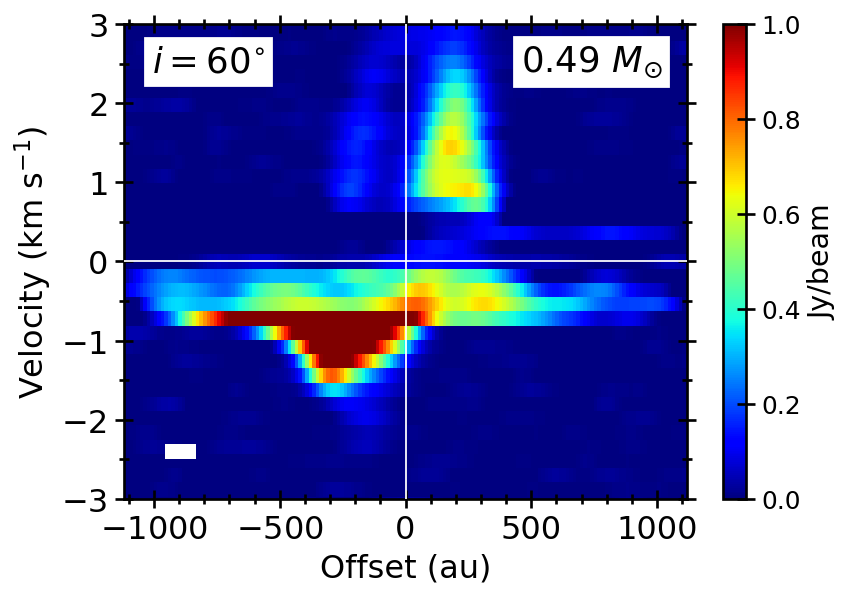}{0.38\textwidth}{(d) Along the black dashed line in Figure \ref{fig:60mom}(b).}
}
\caption{Same as Figure \ref{fig:fidpv} but with the inclination angle of $i=60\arcdeg$.
The diagrams are generated along the black solid and dashed lines in Figure~\ref{fig:60mom}.
\label{fig:60pv}}
\end{figure}

Figure \ref{fig:60pv} shows the PV diagrams along the outflow axis (panels (a) and (b)) and along a cut perpendicular to the outflow at the declination offset for $i=60\arcdeg$ (panels (c) and (d)).
The PV diagram (Figure~\ref{fig:60pv} (a) and (b)) along the outflow axis or the black solid line in Figure~\ref{fig:60mom} exhibits a velocity gradient from the positive offset (blueshifted) to the negative offset (redshifted), which results from the vertical velocity $V_z$.
This gradient is more pronounced than in the PV diagram for $i=85\arcdeg$ (Figures \ref{fig:fidpv}(a) and \ref{fig:fidpv}(b)) in both cases, with and without the interferometric effect.
This is because a lower inclination angle increases the contribution of $V_z$ to the line-of-sight velocity.

Figures \ref{fig:60pv}(c) and \ref{fig:60pv}(d) present the PV diagrams along the cut perpendicular to the outflow or along the the black dashed line in Figure~\ref{fig:60mom}.
The declination offset is smaller than in the $i=85\arcdeg$ case to ensure that this cut intersects the outflow axis at an offset of 540~au along the outflow axis.
In this figure, the most redshifted emission appears at an offset of approximately $+$200~au, while the most blueshifted emission is located around an offset of approximately $-200$~au.
This pattern is observed with and without the interferometric effect.
This velocity gradient is attributed to outflow rotation.
However, identifying rotational velocity from this PV diagram requires knowledge of the outflow direction, as the velocity gradient must be evaluated with respect to it.

\begin{figure}[htbp]
\gridline{
\fig{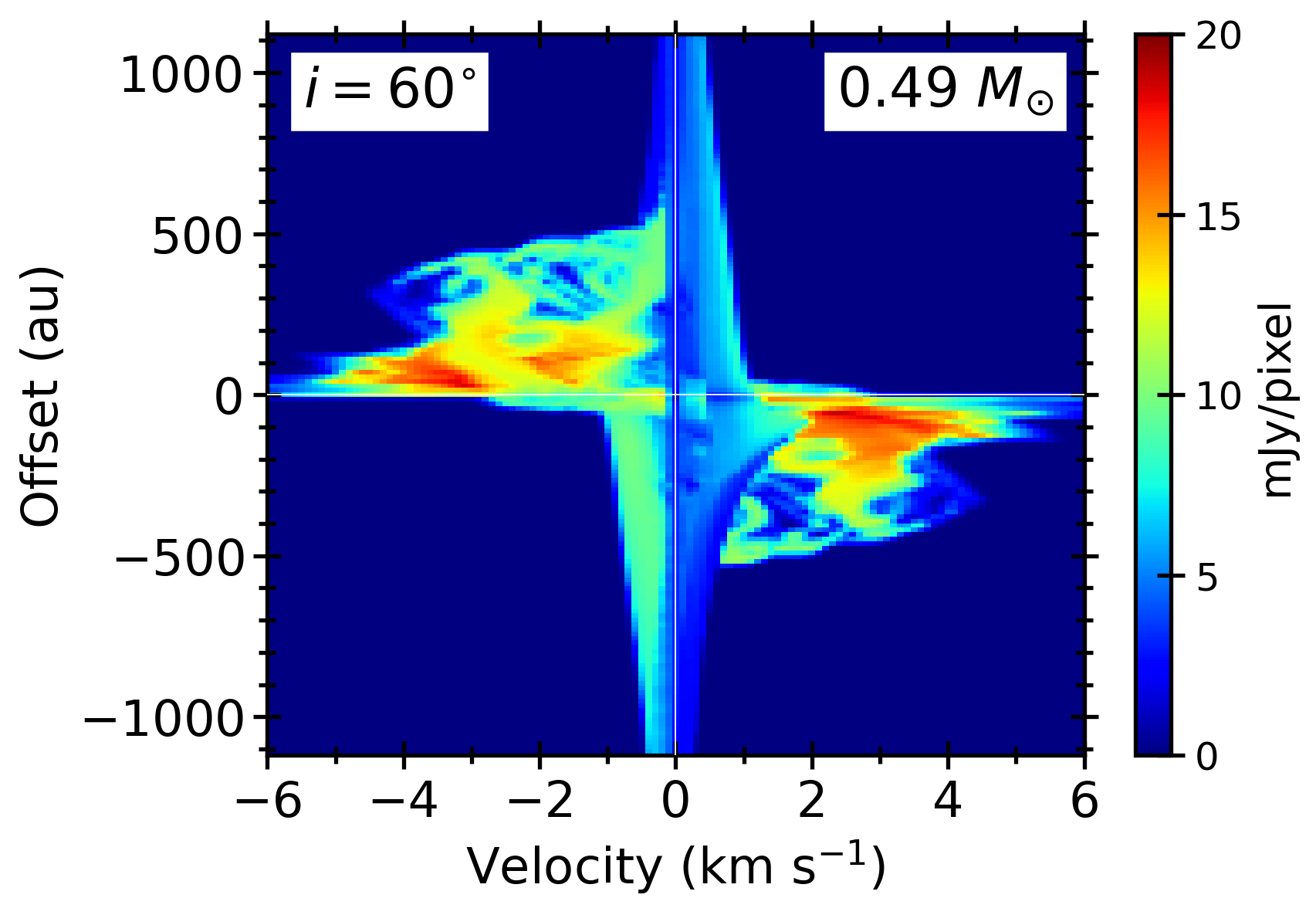}{0.42\textwidth}{(a) Along the gray solid line in Figure \ref{fig:60mom}(a).}
\fig{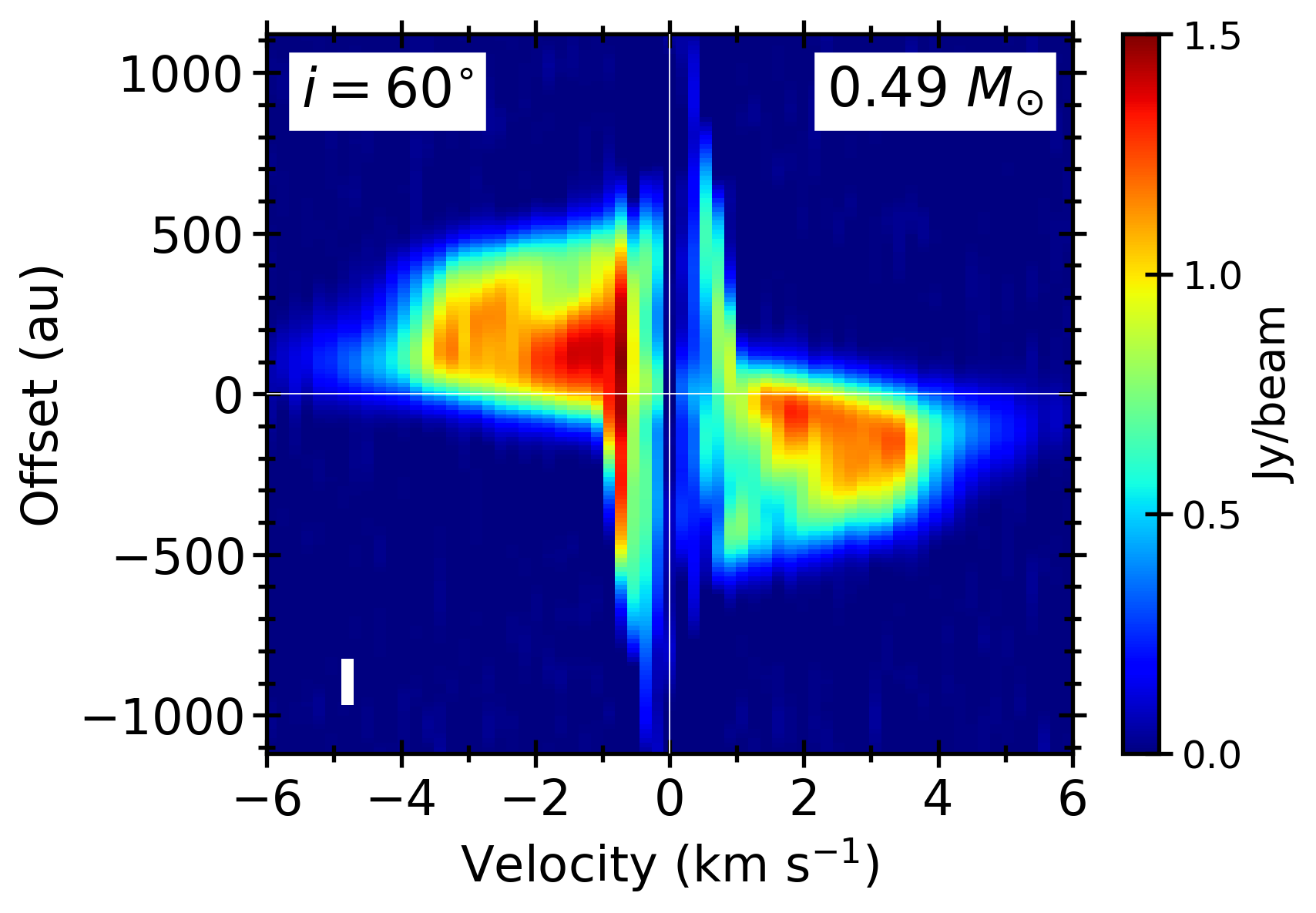}{0.42\textwidth}{(b) Along the gray solid line in Figure \ref{fig:60mom}(b).}
}
\gridline{
\fig{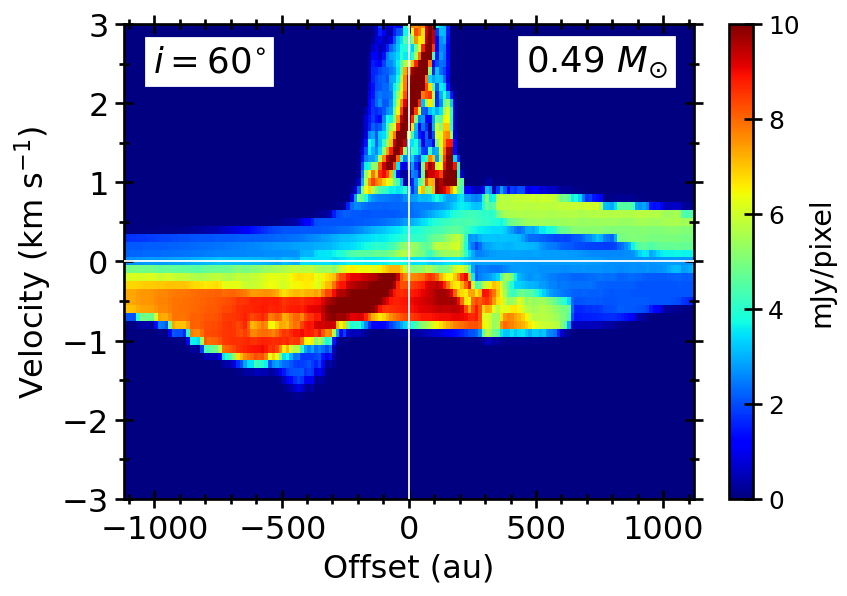}{0.38\textwidth}{(c) Along the gray dashed line in Figure \ref{fig:60mom}(a).}
\fig{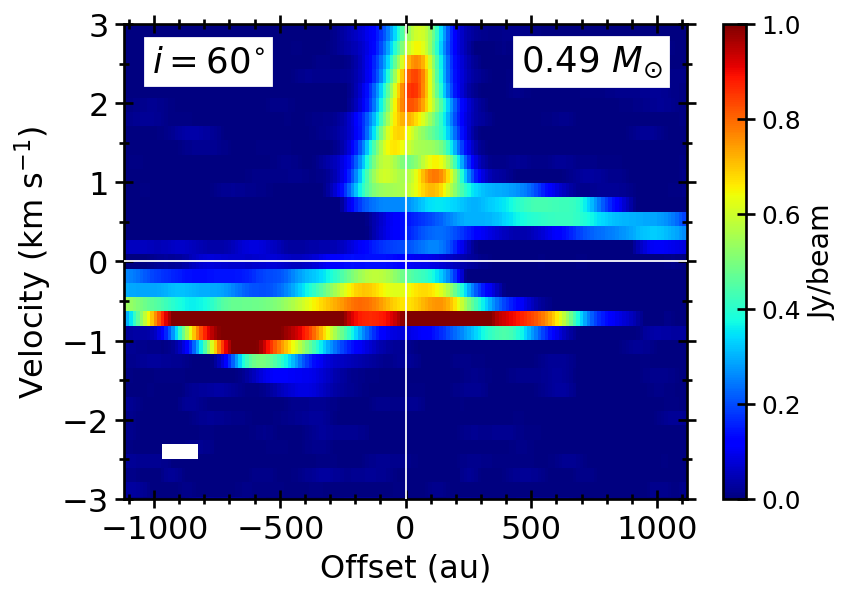}{0.38\textwidth}{(d) Along the gray dashed line in Figure \ref{fig:60mom}(b).}
}
\caption{Same as Figure~\ref{fig:60pv}, but generated along differently oriented cuts (gray solid and dashed lines in Figure~\ref{fig:60mom}).
\label{fig:60pv2}}
\end{figure}

It is therefore valuable to examine how the PV diagram changes when the cutting direction differs from the ``correct" direction used above. 
In the moment 1 map (Figure \ref{fig:60mom}), the most prominent velocity gradient could be interpreted as the outflow (propagation) direction if no additional information available.
To investigate this effect, the two cuts for the PV diagrams are tilted by $40\arcdeg$ as shown with the gray solid and dashed lines in Figure \ref{fig:60mom}. 
Figure \ref{fig:60pv2} shows the PV diagrams along the tilted cuts. 
The diagrams along the gray solid cuts (Figures \ref{fig:60pv2}(a) and \ref{fig:60pv2}(b)) show a clearer velocity gradient from the positive offset (blueshifted velocity) to the negative offset (redshifted velocity) compared to the diagrams along the black solid (i.e., ``correct") cuts (Figures \ref{fig:60pv}(a) and \ref{fig:60pv}(b)).
This result thus does not indicate that the cutting direction is incorrect. 
The diagrams along the gray dashed cuts (Figures~\ref{fig:60pv2}(c) and \ref{fig:60pv2}(d)) show that the redshifted emission at $>1~\kms$ is concentrated around the zero offset. 
Although the blueshifted emission at $<-1~\kms$ appears around an offset of $\sim -600$~au because of outflow rotation, this component is difficult to distinguish from the low-velocity extended component, particularly when it is compared to its counterparts in the diagrams along the black dashed (i.e., ``correct") cuts (Figures~\ref{fig:60pv}(c) and \ref{fig:60pv}(d)). 
The redshifted and blueshifted components in Figures~\ref{fig:60pv2}(c) and \ref{fig:60pv2}(d) do not produce a clear velocity gradient in the PV diagrams along the direction perpendicular to the misinterpreted outflow axis. 
In other words, an incorrect outflow axis hinders the observational identification of outflow rotation in both the moment map and the PV diagram.


\section{Discussion} \label{sec:discussion}
Our analysis suggests that a low inclination angle can hinder the identification of outflow rotation by distorting the spatial distribution of observed emission and velocity.
In this section, we discuss the mechanism behind this distortion and examine complex observational results that highlight the challenges in determining the outflow axis.

\subsection{Difference between Outflow and Infall} \label{sec:outin}
\begin{figure}[htbp]
\fig{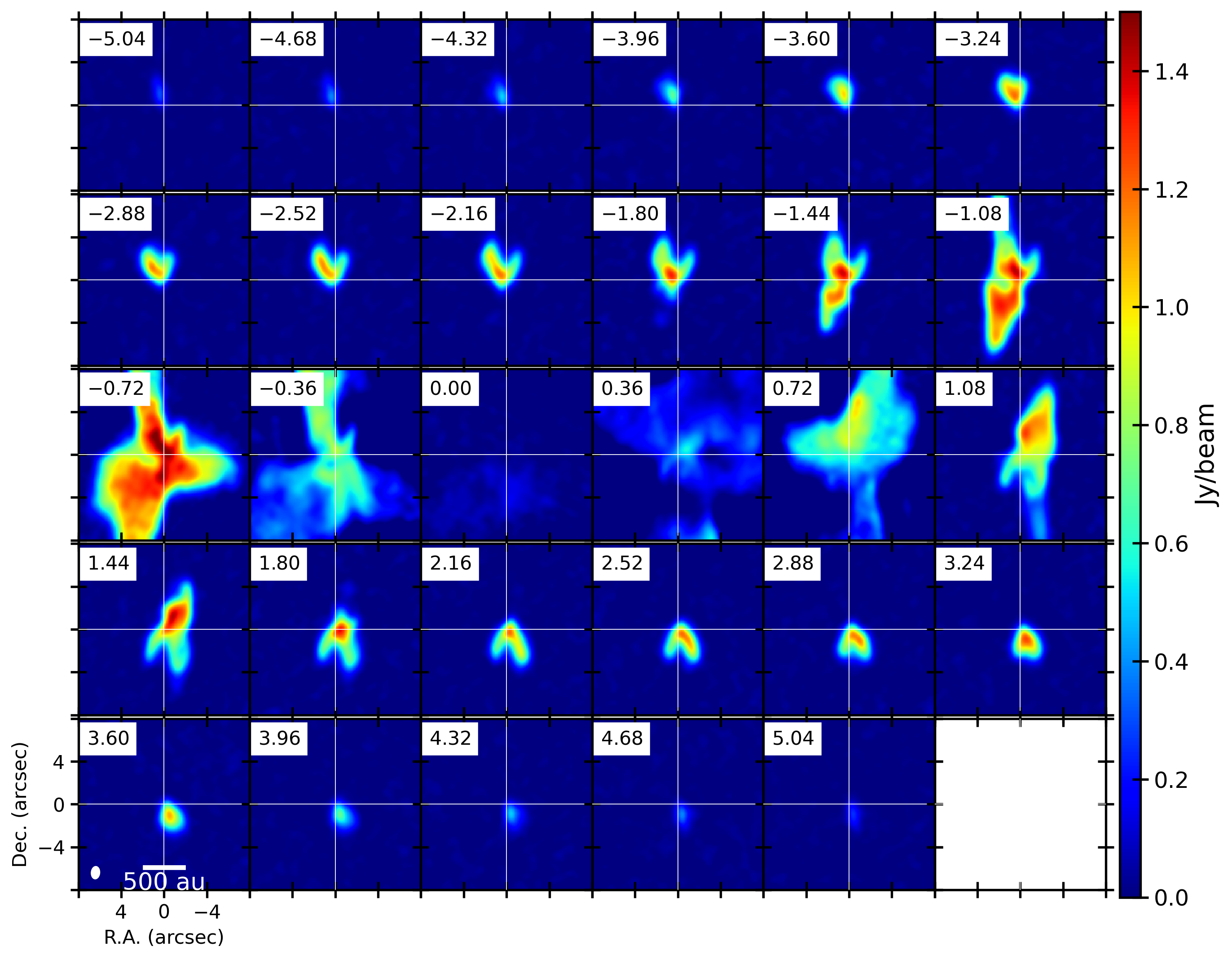}{0.99\textwidth}{}
\caption{
Channel maps of the snapshot for the CB 26 model with an inclination angle of $i=60\arcdeg$. 
The number in the top left corner indicates the line-of-sight velocity of each channel in units of $\kms$. 
The maps are displayed at intervals of two channels to present the entire velocity structure with moderately large panels.
The data cube used to generate these channel maps has a velocity resolution of $0.18~\kms$ (Section \ref{sec:settings}).
\label{fig:chan}}
\end{figure}

The two velocity gradients in the moment 1 map for  $i=60\arcdeg$ can be decomposed using channel maps, as shown in Figure~\ref{fig:chan}, which includes the interferometric effect. 
The high-velocity component at $|V|\gtrsim 2~\kms$ is primarily elongated from the central stellar position toward the upper left in the blueshifted channels and toward the lower right in the redshifted channels.
As a result, even when using both the moment 0 map and channel maps, this elongated direction could be misinterpreted as the outflow direction if one expects jet-like emission at high velocities as reported in previous CO line observations of protostellar outflows \citep[e.g.,][]{hira10, aso18}.

\begin{figure}[htbp]
\fig{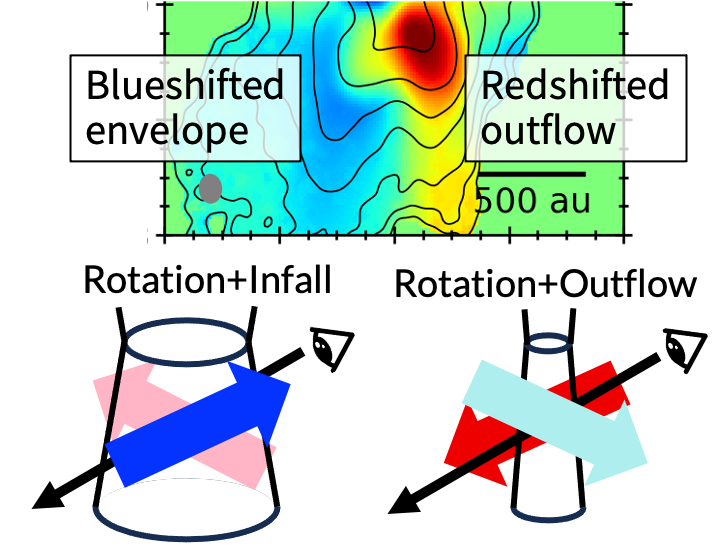}{0.49\textwidth}{}
\caption{
Schematic view of the difference between rotation plus outflow motion and  rotation plus infall motion. 
Darker colors indicate higher line-of-sight velocities, while the fainter color represent line-of-sight velocities closer to the systemic velocity. This difference results in redshifted emission, originating from the outflow, appearing on the right side, and the blueshifted emission, originating from the envelope on the left side, in the moment 1 map.
\label{fig:schem}}
\end{figure}

In contrast to the high-velocity component, the low-velocity component at $0.5~\kms \lesssim |V| \lesssim 1~\kms$ is more elongated toward the lower left in the blueshifted channels and toward the upper right in the redshifted channels (Figure~\ref{fig:chan}).
Emission is largely resolved out in the synthetic interferometric observation at $|V|\lesssim 0.5~\kms$. 
The direction of elongation and the velocity gradient of the low-velocity component are diagonal to those of the high-velocity component. 
This difference can be attributed to the difference between the outflow and the infalling envelope in the same system, as summarized in Figure \ref{fig:schem}. 
In this schematic view, both the rotational velocity and the outflow velocity (directed downward in this figure) are redshifted on the right side when viewed by the observer.
On the left side, the rotational velocity is blueshifted, while the vertical outflow velocity remains redshifted.
As a result, the emission on the right side is observed at a higher line-of-sight velocity than that on the left side.
The opposite combination occurs in the case of rotation combined with vertical infall (directed upward in this figure), corresponding to an infalling and rotating envelope.
In this case, the emission on the left side appears at a higher velocity than that on the right side.
If the combined velocity is close to the systemic velocity, the emission may be resolved out in an interferometric observation.
In addition, infalling gas is expected to be located at larger cylindrical radii than outflowing gas.
These effects may lead to an asymmetric structure in both the moment 0 and moment 1 maps.


\subsection{Inclination Effects during Protostellar Evolution} \label{sec:evo}
In Sections \ref{sec:anares} and \ref{sec:outin}, to compare the observational example of the CB 26 outflow with our simulation results, we analyzed and discussed a specific snapshot at two inclination angles $i=60\arcdeg$ and $85\arcdeg$. 
Meanwhile, our MHD simulation covers the evolutionary stages from $M_*\sim 0$ to $\sim 0.5~\Msun$, where the central protostellar mass $M_*$ increases as time goes on through mass accretion into the sink cell. Note that it takes $\sim 53,000$ years for $M_*$ to increase from 0 to $\sim 0.5~\Msun$. This evolutionary coverage allows us to examine the difficulty in identifying outflow rotation at different evolutionary stages as well as at different inclination angles. 
The channel maps in Figure~\ref{fig:chan} show that the outflow emission primarily appears in the high-velocity range, while the low-velocity range includes more emission from an associated infalling envelope. 
Therefore, the use of the high-velocity component is a reasonable approach to determining the outflow axis, as it minimizes contamination from envelope emission.
To further reduce contamination, we created moment 0 and moment 1 maps with the interferometric effect by integrating the intensity over a velocity range of $|V|=1.44-5.04~\kms$. 

\begin{figure}[htbp]
\fig{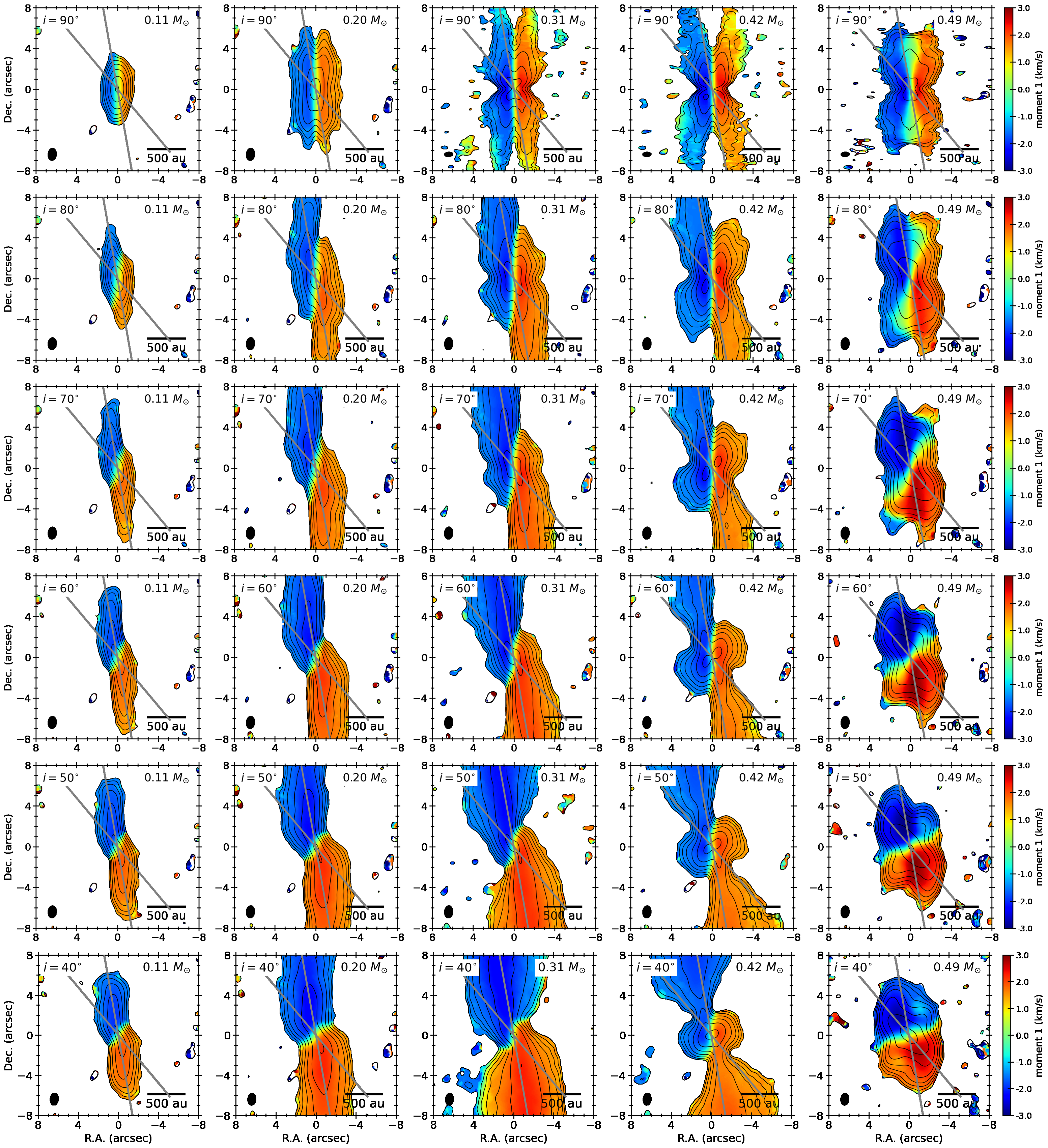}{0.90\textwidth}{}
\caption{Moment 0 and 1 maps with different evolutionary stages (different protostellar mass $M_*$) and different inclination angles. The intensities are integrated over a velocity range of $|V|=1.44$ to $5.04~\kms$ to suppress the emission of the infalling envelope. The gray lines denote the position angles of $10\arcdeg$ and $40\arcdeg$. The contour levels are $3,6,12,24,...\times 8~\mJB$. These evolutionary stages are the same as those in \citet{as.ma20} (the fifth column is not the snapshot for the CB 26 model used in Sections \ref{sec:anares} and \ref{sec:outin}).
\label{fig:evo}}
\end{figure}

Figure \ref{fig:evo} shows the moment maps with this velocity range at the evolutionary stages of $M_* \sim 0.1, 0.2, 0.3, 0.4$, and $0.5~\Msun$ and the inclination angles of $i= 90\arcdeg, 80\arcdeg, 70\arcdeg, 60\arcdeg, 50\arcdeg$, and $40\arcdeg$. 
The evolutionary stages correspond to the five representative stages in  \citet{as.ma20}. 
All pairs of $M_*$ and $i$ exhibit an elongated structure, possibly with an S-shape, except for some pairs with the largest $M_*$ and lower $i$. 
The shorter emission sizes at the most evolved stage are due to the sensitivity limit, not the intrinsic outflow length, as the outflow is less dense at this stage than at earlier stages.
In most cases, this elongated direction is tilted counterclockwise by $10\arcdeg -20\arcdeg$ from the vertical direction corresponding to the actual outflow axis. 
For $i\leq 60\arcdeg$, the overall velocity gradient also aligns primarily with the elongated direction. 
This combination of elongation and velocity gradient could lead to a slightly tilted and incorrect outflow axis. 
In some cases at $M_*\sim 0.3$ or $0.4~\Msun$ (and $i\leq 70\arcdeg$), an additional velocity gradient appears perpendicular to the elongated direction within the central $\lesssim 200$-au region. 
This velocity gradient results from contamination by the inner envelope (Figure \ref{fig:schem}).
This feature can be misinterpreted as a result of rotation around the incorrect outflow axis, thereby reinforcing the misidentified axis.
Such misinterpretation does not occur in panels with $i\sim 90\arcdeg$ because the velocity gradient remains horizontal over a wide range of declination offsets, which is a clear signature of outflow rotation.
Cases with $i=80\arcdeg$ may lead to misinterpretation at younger stages but not at more evolved stages.


In summary, the synthetic observations of our MHD simulation suggest that misinterpretation of the outflow axis, leading to non-detection of outflow rotation, may occur when the inclination angle is $i \lesssim 70\arcdeg - 80\arcdeg$. 
This angle range indicates that due to the inclination effect, the rotation of  $(\int _{-i}^{i} \int _{-\pi}^{\pi} \sin \theta d\phi d\theta / 4\pi=) 66\%- 83\%$ i.e., $\sim 2/3$ to $\sim 4/5$ of protostellar outflows remains undetectable, assuming a random configuration in 3D space relative to the observer.
Furthermore, this effect cannot be mitigated by higher angular resolution or increased sensitivity, because our analysis shows that even in the absence of interferometric effects, the same result holds (Section \ref{sec:anares}).


\subsection{Complex Outflow Axes in Observational Studies} \label{sec:complex}
One may wonder how challenging it is to determine the outflow axis in actual observational studies.
\citet{bjer16} observed the outflow of a protostellar system, TMC-1A, at an inclination angle of  $i=50-60\arcdeg$, using the CO $J=2-1$ line, and identified its rotation.
In their observation, the blueshifted emission is detected only on the eastern side of the outflow axis, extending over a length of $\sim 100$~au.
The authors attribute this difference between the eastern and western sides to a combination of vertical and rotational velocities in the outflow.
This idea is consistent with the discussion in Section \ref{sec:outin}, and similarly, concentrated emission also appears in our channel maps (Figure \ref{fig:chan}).
The outflow axis is primarily determined based on observations at a spatial scale of thousands of au, and the associated Keplerian disk and infalling envelope aid in determining the outflow axis \citep[see references in][]{bjer16}, which reveals that the outflow is perpendicular to the disk plane.
This example of a rotating outflow demonstrates that emission appearing on one side of the outflow axis in channel maps or in distorted moment maps can be correctly interpreted as a result of outflow rotation, when we have other observational studies on the associated disk/envelope as well as on a 1000-au-scale outflow, which provide that the outflow is perpendicular to the disk.

On the other hand, previous observational studies have shown more complex protostellar systems, where the disk/envelope or larger-scale outflow may not aid in determining the outflow axis.  
For example, \citet{mats19} observed a protostellar system, MMS 5, within the Orion Molecular Cloud 3 and detected a 5000-au-sized outflow in the CO \(J=2-1\) line and a 1000-au-sized jet in the SiO \(J=5-4\) line.  
Their observations revealed that the outflow direction on the plane of the sky is \(\mathrm{P.A.} = 79\arcdeg\), whereas the jet direction is \(\mathrm{P.A.} = 96\arcdeg\).  
These two directions differ from each other and are not perpendicular to the elongated direction of the dust continuum emission (\(\mathrm{P.A.} = 144\arcdeg\)), which is often used as an indicator of the disk orientation.  
In addition, their channel maps in the two lines show that the blueshifted and redshifted lobes of the outflow point in different directions, as do the jet lobes.  

A severer bending is reported in the CO outflow of a protostellar system, SMM4B, in the Serpens Main region \citep{aso18}.  
The blueshifted lobe has a direction of \(\mathrm{P.A.} \sim 0\arcdeg\), whereas the redshifted lobe has a direction of \(\mathrm{P.A.} = 150\arcdeg\), from 3000-au down to \(\lesssim 400\)-au scales.  
The associated continuum emission is elongated in a different direction of \(\mathrm{P.A.} = 95\arcdeg\).  
\citet{chin16} observed two CO outflows on the thousands-au scale in a protobinary system, NGC 1333 IRAS 4A, in the Perseus molecular cloud.  
The authors suggest that the redshifted lobes of the two outflows have merged into a single lobe due to magnetic effects, while the blueshifted lobes do not overlap with each other.  

Other observational studies report that the direction of a protostellar outflow has changed over time in a single protostellar system, IRAS 15398-3359, due to turbulence in its natal core \citep{okod21, sai24}.  
In addition, outflow precession produces an S-shaped outflow, which may appear similar to the S-shape found in some panels of Figure \ref{fig:evo}.  
These complicated outflows highlight the difficulty in determining the outflow axis.  
For these reasons, we suggest that ideal targets for detecting outflow rotation are limited to highly inclined outflows with a well-defined bipolar configuration in current observational studies and are thus significantly fewer than expected from theoretical studies.

\subsection{Effect of High Vertical Velocity} \label{sec:vertical}
A lower inclination angle causes the outflow vertical (rotational) velocity to contribute more (less) to the line-of-sight velocity.  
A higher intrinsic ratio between vertical and rotational velocities would have the same effect.  
Thus, the combination of a high $V_z$ and inclination effects would raise the inclination threshold for detecting outflow rotation, which is  $i=70\arcdeg-80\arcdeg$  in our analysis, and could therefore obscure rotation in a greater number of observed outflows.  

The vertical and rotational velocities in the snapshot of the CB 26 model are consistent with observations (Table~\ref{tab:comp}).  
The vertical (outflow) velocity of $\lesssim 10~\kms$ is also a typical value in survey observations \citep{step19}.  
These observational results support the velocity ratio in our model.  
Meanwhile, vertical (outflow) velocities of $\gtrsim 30~\kms$ have also been found in observational studies \citep[e.g.,][]{hira10, aso18, omur24}. 
In numerical studies, such a fast outflow could be suppressed by a sink cell  because the faster parts of an outflow are launched from smaller radii within the sink region.  
In other words, adopting a smaller or no sink cell will produce a gas flow launched from smaller radii, which will add more (linear) momentum but little angular momentum to the outflow.  
This component could thus increase the vertical velocity relative to the rotational velocity, further obscuring outflow rotation in observations.  
The impact of this effect must be quantitatively evaluated using a future MHD simulation with a smaller sink cell, in order to determine how much outflow rotation is obscured in total by the combination of the inclination and high-$V_z$ effects.


\section{Conclusions} \label{sec:conc}
We examined observational signatures of outflow rotation using an MHD simulation of protostellar evolution.  
The temperature and intensity at each snapshot of the simulation were calculated through radiative transfer using RADMC-3D.  
The interferometric effect was incorporated through the CASA {\it simobserve}. 
Using the image cube obtained after these steps, we created moment 0 and 1 maps, as well as position-velocity diagrams along a cut perpendicular to the outflow axis with an offset from the midplane.  
We then analyzed velocity gradients and emission morphology in these figures as signatures of outflow rotation.  
The main results are summarized below.  

\begin{enumerate}
\item 
A snapshot in our simulation shows an outflow lobe with a length, velocity, and mass similar to an observational example of the outflow in a protostellar system, CB 26.  
With the same inclination angle as the CB 26 outflow ($i=85\arcdeg$), signatures of outflow rotation can be identified in the moment maps and PV diagrams.  
The moment 0 map shows the main structure elongated in the outflow direction, with a neck at the midplane.  
The moment 1 map reveals a primary velocity gradient in the direction perpendicular to the outflow axis, which is also observed in the PV diagram across the lobe.  

\item 
The effects of a different inclination angle are investigated using the same snapshot but with a smaller inclination angle of $i=60\arcdeg$.  
The moment 1 map and the PV diagram show a velocity gradient perpendicular to the outflow axis.  
Meanwhile, the moment 0 and 1 maps are distorted compared to those at $i=85\arcdeg$ due to the combination of vertical and rotational velocities.  
In particular, the moment 1 map could lead to an incorrect determination of the outflow axis if the true axis is not known.  
With this incorrect axis, the PV diagram across the lobe appears to show no velocity gradient (i.e., no rotation signature).  
The PV diagram along the incorrect axis does not provide evidence that the chosen axis is incorrect.  
The velocity combination distorts the moment maps also in the infalling envelope. The distortion direction is different from that in the outflow because the vertical motion of the envelope is opposite to that of the outflow.  

\item 
The misinterpretation of the outflow axis could occur when the inclination angle is $i\lesssim 70\arcdeg -80\arcdeg$ across all evolutionary stages simulated ($M_*\sim 0$ to $\sim 0.5~\Msun$). 
This indicates that 66\% to 83\% of rotating outflows may be obscured by the inclination effect.  
The complex structures observed in outflows further highlight the difficulty in determining the outflow axis, including bending, misalignment with a jet, overlap, and precession.  
Our results suggest the inclination effect as a possible explanation for the observational scarcity of outflow rotation.  
While ideal targets for studying outflow rotation are outflows that are nearly parallel to the plane of the sky and exhibit symmetric morphology relative to their axis, the selection bias introduced by the inclination effect must be considered in statistical and population studies on outflow rotation.  
\end{enumerate}

\begin{acknowledgments}
We acknowledge the use of the gmunu HPC cluster of KASI for the radiative transfer calculations.
This research used the computational resources of the HPCI system provided by the Cyber Science Center at Tohoku University and the Cybermedia Center at Osaka University (Project ID:hp230035, hp240010).
Simulations reported in this paper were also performed by 2023 and 2024 Koubo Kadai on Earth Simulator (NEC SX-ACE) at JAMSTEC. 
This work was supported by a NAOJ ALMA Scientific Research grant (No. 2022-22B). 
This work was also supported by JSPS KAKENHI Grant Number JP25K07369. 
\end{acknowledgments}

\begin{contribution}

YA analyzed the synthetic observational data. MM provided the numerical simulation data for the synthetic observation. Both authors contributed equally to the other aspects of this work.


\end{contribution}

%

\software{astropy \citep{astr13, astr18}, CASA \citep{casa22}, numpy \citep{harr20}}





\begin{thebibliography}{}

\bibitem[Arce \& Sargent(2006)]{ar.sa06} Arce, H.~G. \& Sargent, A.~I.\ 2006, \apj, 646, 1070. doi:10.1086/505104
\bibitem[Aso et al.(2018)]{aso18} Aso, Y., Hirano, N., Aikawa, Y., et al.\ 2018, \apj, 863, 19. doi:10.3847/1538-4357/aacf9b
\bibitem[Aso \& Machida(2020)]{as.ma20} Aso, Y. \& Machida, M.~N.\ 2020, \apj, 905, 174. doi:10.3847/1538-4357/abc6fc
\bibitem[Astropy Collaboration et al.(2018)]{astr18} Astropy Collaboration, Price-Whelan, A.~M., Sip{\H{o}}cz, B.~M., et al.\ 2018, \aj, 156, 123. doi:10.3847/1538-3881/aabc4f
\bibitem[Astropy Collaboration et al.(2013)]{astr13} Astropy Collaboration, Robitaille, T.~P., Tollerud, E.~J., et al.\ 2013, \aap, 558, A33. doi:10.1051/0004-6361/201322068
\bibitem[Bjerkeli et al.(2016)]{bjer16} Bjerkeli, P., van der Wiel, M.~H.~D., Harsono, D., et al.\ 2016, \nat, 540, 406. doi:10.1038/nature20600
\bibitem[Blandford \& Payne(1982)]{bl.pa82} Blandford, R.~D. \& Payne, D.~G.\ 1982, \mnras, 199, 883. doi:10.1093/mnras/199.4.883
\bibitem[CASA Team et al.(2022)]{casa22} CASA Team, Bean, B., Bhatnagar, S., et al.\ 2022, \pasp, 134, 1041, 114501. doi:10.1088/1538-3873/ac9642
778d776
\bibitem[Ching et al.(2016)]{chin16} Ching, T.-C., Lai, S.-P., Zhang, Q., et al.\ 2016, \apj, 819, 159. doi:10.3847/0004-637X/819/2/159
\bibitem[de Valon et al.(2022)]{deva22} de Valon, A., Dougados, C., Cabrit, S., et al.\ 2022, \aap, Modeling the CO outflow in DG Tauri B: Swept-up shells versus perturbed MHD disk wind, 668, A78. doi:10.1051/0004-6361/202141316
\bibitem[Dullemond et al.(2012)]{dull12} Dullemond, C.~P., Juhasz, A., Pohl, A., et al.\ 2012, Astrophysics Source Code Library. ascl:1202.015
\bibitem[G{\'o}mez-Ruiz et al.(2019)]{gome19} G{\'o}mez-Ruiz, A.~I., Gusdorf, A., Leurini, S., et al.\ 2019, \aap, 629, A77. doi:10.1051/0004-6361/201424156
789d788
\bibitem[Goodman et al.(1993)]{good93} Goodman, A.~A., Benson, P.~J., Fuller, G.~A., et al.\ 1993, \apj, 406, 528. doi:10.1086/172465
\bibitem[Harris et al.(2020)]{harr20} Harris, C.~R., Millman, K.~J., van der Walt, S.~J., et al.\ 2020, \nat, 585, 357. doi:10.1038/s41586-020-2649-2
\bibitem[Hirano et al.(2010)]{hira10} Hirano, N., Ho, P.~P.~T., Liu, S.-Y., et al.\ 2010, \apj, 717, 58. doi:10.1088/0004-637X/717/1/58
\bibitem[Hirota et al.(2017)]{hiro17} Hirota, T., Machida, M.~N., Matsushita, Y., et al.\ 2017, Nature Astronomy, 1, 0146. doi:10.1038/s41550-017-0146
\bibitem[Kauffmann et al.(2008)]{kauf08} Kauffmann, J., Bertoldi, F., Bourke, T.~L., et al.\ 2008, \aap, 487, 993. doi:10.1051/0004-6361:200809481
\bibitem[Lacy et al.(1994)]{lacy94} Lacy, J.~H., Knacke, R., Geballe, T.~R., et al.\ 1994, \apjl, 428, L69. doi:10.1086/187395
\bibitem[Launhardt et al.(2023)]{laun23} Launhardt, R., Pavlyuchenkov, Y.~N., Akimkin, V.~V., et al.\ 2023, \aap, 678, A135. doi:10.1051/0004-6361/202347483
\bibitem[Lee et al.(2017)]{lee17} Lee, C.-F., Ho, P.~T.~P., Li, Z.-Y., et al.\ 2017, Nature Astronomy, A rotating protostellar jet launched from the innermost disk of HH 212, 1, 0152. doi:10.1038/s41550-017-0152
\bibitem[L{\'o}pez-V{\'a}zquez et al.(2023)]{lope23} L{\'o}pez-V{\'a}zquez, J.~A., Zapata, L.~A., \& Lee, C.-F.\ 2023, \apj, 944, 63. doi:10.3847/1538-4357/acb439
\bibitem[L{\'o}pez-V{\'a}zquez et al.(2020)]{lope20} L{\'o}pez-V{\'a}zquez, J.~A., Zapata, L.~A., Lizano, S., et al.\ 2020, \apj, ALMA Observations and Modeling of the Rotating Outflow in Orion Source I, 904, 2, 158. doi:10.3847/1538-4357/abbe24
\bibitem[Machida et al.(2004)]{Machida2004} Machida, M.~N., Tomisaka, K., \& Matsumoto, T.\ 2004, \mnras, 348, L1. doi:10.1111/j.1365-2966.2004.07402.x
\bibitem[Machida et al.(2006)]{Machida2006} Machida, M.~N., Inutsuka, S.-. ichiro ., \& Matsumoto, T.\ 2006, \apjl, 647, L151. doi:10.1086/507179
\bibitem[Machida et al.(2014)]{Machida2014} Machida, M.~N., Inutsuka, S.-. ichiro ., \& Matsumoto, T.\ 2014, \mnras, 438, 2278. doi:10.1093/mnras/stt2343
\bibitem[Machida et al.(2016)]{Machida2016} Machida, M.~N., Matsumoto, T., \& Inutsuka, S.-. ichiro .\ 2016, \mnras, 463, 4246. doi:10.1093/mnras/stw2256
\bibitem[Machida et al.(2008)]{mach08} Machida, M.~N., Inutsuka, S.-. ichiro ., \& Matsumoto, T.\ 2008, \apj, 676, 1088. doi:10.1086/528364
\bibitem[Machida \& Basu(2024)]{ma.ba24} Machida, M.~N. \& Basu, S.\ 2024, \apj, 970, 41. doi:10.3847/1538-4357/ad4997
\bibitem[Machida \& Hosokawa(2013)]{ma.ho13} Machida, M.~N. \& Hosokawa, T.\ 2013, \mnras, 431, 2, 1719. doi:10.1093/mnras/stt291
\bibitem[Matsushita et al.(2021)]{mats21} Matsushita, Y., Takahashi, S., Ishii, S., et al.\ 2021, \apj, 916, 23. doi:10.3847/1538-4357/ac069f
\bibitem[Matsushita et al.(2019)]{mats19} Matsushita, Y., Takahashi, S., Machida, M.~N., et al.\ 2019, \apj, 871, 221. doi:10.3847/1538-4357/aaf1b6
\bibitem[Okoda et al.(2021)]{okod21} Okoda, Y., Oya, Y., Francis, L., et al.\ 2021, \apj, 910, 11. doi:10.3847/1538-4357/abddb1
\bibitem[Omura et al.(2024)]{omur24} Omura, M., Tokuda, K., \& Machida, M.~N.\ 2024, \apj, 963, 72. doi:10.3847/1538-4357/ad19ce
\bibitem[Stephens et al.(2019)]{step19} Stephens, I.~W., Bourke, T.~L., Dunham, M.~M., et al.\ 2019, \apjs, 245, 21. doi:10.3847/1538-4365/ab5181
\bibitem[Pudritz \& Norman(1986)]{pu.no86} Pudritz, R.~E. \& Norman, C.~A.\ 1986, \apj, 301, 571. doi:10.1086/163924
\bibitem[Sai et al.(2024)]{sai24} Sai, J., Yen, H.-W., Machida, M.~N., et al.\ 2024, \apj, 966, 192. doi:10.3847/1538-4357/ad34b7
\bibitem[Tomida et al.(2017)]{tomi17} Tomida, K., Machida, M.~N., Hosokawa, T., et al.\ 2017, \apjl, 835, L11. doi:10.3847/2041-8213/835/1/L11
\bibitem[Zapata et al.(2010)]{zapa10} Zapata, L.~A., Schmid-Burgk, J., Muders, D., et al.\ 2010, \aap, A rotating molecular jet in Orion, 510, A2. doi:10.1051/0004-6361/200810245

\end{thebibliography}



\end{document}